# Differentially Private $k$-Means with Constant Multiplicative Error

Haim Kaplan*     Uri Stemmer†

July 16, 2018

**Abstract**

We design new differentially private algorithms for the Euclidean $k$-means problem, both in the *centralized model* and in the *local model* of differential privacy. In both models, our algorithms achieve significantly improved error guarantees than the previous state-of-the-art. In addition, in the local model, our algorithm significantly reduces the number of interaction rounds.

Although the problem has been widely studied in the context of differential privacy, all of the existing constructions achieve only super constant approximation factors. We present—for the first time—efficient private algorithms for the problem with *constant* multiplicative error. Furthermore, we show how to modify our algorithms so they compute *private corsets* for $k$-means clustering in both models.

## 1 Introduction

Clustering, and in particular center based clustering, are central problems in unsupervised learning. Several cost objectives have been intensively studied for center based clustering, such as minimizing the sum or the maximum of the distances of the input points to the centers. Most often the data is embedded in Euclidean space and the distances that we work with are Euclidean distances. In particular, one of the most studied center based clustering problem is the *Euclidean $k$-means problem*. In this problem we are given a set of $n$ input points in $\mathbb{R}^d$ and our goal is to find $k$ centers that minimize the sum of squared distances between each input point to its nearest center.[1] When privacy is not a concern one usually solves this problem by running Lloyd's algorithm [34] initialized by $k$-means++ [6]. This produces $k$-centers of cost that is no worse than $O(\log k)$ times the cost of the optimal solution and its performance in practice is typically better.

The huge applicability of $k$-means clustering, together with the increasing awareness and demand for user privacy, motivated the study of *privacy preserving $k$-means algorithms*. It is especially desirable to achieve *differential privacy* [20], a privacy notion which has been widely adopted by the academic community as well as big corporations like Google, Apple, and Microsoft. Indeed, constructions of differentially private $k$-means algorithms have received a lot of attention over the last 14 years [14, 38, 22, 25, 37, 45, 41, 43, 24, 8, 39, 29]. In this work we design new differentially private $k$-means algorithms, both for the centralized model (where a trusted curator collects the sensitive information and analyzes it with differential privacy) and for the local model (where each respondent randomizes her answers to the data curator to protect her privacy). In both models, our algorithms offer significant improvements over the previous state-of-the-art.

---

*School of Computer Science, Tel Aviv University and Google. `haimk@post.tau.ac.il`
†Dept. of Computer Science and Applied Math, Weizmann Institute of Science. `u@uri.co.il`

[1]The sum of squares is nice to work with since we do not have to compute square roots. Furthermore, for a given cluster its center of mass is the minimizer of the sum of the squared distances. These properties make $k$-means to be the favorite for center based clustering.

Before describing our new results, we define our setting more precisely. Consider an input database $S = (x_1, \ldots, x_n) \in (\mathbb{R}^d)^n$ containing $n$ points in $\mathbb{R}^d$, where every point $x_i \in S$ is the (sensitive) information of one individual. The goal is to identify a set of $k$ *centers* $C = \{c_1, \ldots, c_k\}$ in $\mathbb{R}^d$ approximately minimizing the following quantity, referred to as the *cost* of the centers

$$\operatorname{cost}_S(C) = \sum_{i=1}^n \min_{j \in [k]} \|x_i - c_j\|_2^2.$$

The privacy requirement is that the output of our algorithm (the set of centers) does not reveal information that is specific to any single individual. Formally,

**Definition 1.1** ([20])**.** *A randomized algorithm $\mathcal{A} : X^n \to Y$ is $(\varepsilon, \delta)$ differentially private if for every two databases $S, S' \in X^n$ that differ in one row, and every set $T \subseteq Y$, we have*

$$\Pr[\mathcal{A}(S) \in T] \leq e^\varepsilon \cdot \Pr[\mathcal{A}(S') \in T] + \delta.$$

Combining the utility and privacy requirements, we are seeking for a computationally efficient differentially private algorithm that identifies a set of $k$ centers $C$ such that w.h.p. $\operatorname{cost}_S(C) \leq \gamma \cdot \operatorname{OPT}_S + \eta$, where $\operatorname{OPT}_S$ is the optimal cost. We want $\gamma$ and $\eta$ to be as small as possible, as a function of the number of input points $n$, the dimension $d$, the number of centers $k$, the failure probability $\beta$, and the privacy parameters $\varepsilon, \delta$.

We remark that a direct consequence of the definition of differential privacy is that, unlike in the non-private literature, every private algorithm for this problem must have additive error $\eta > 0$. In fact, $\eta$ must grow with $\Lambda^2$, where $\Lambda$ bounds the diameter of the space of input points. To see this, consider $k+1$ locations $p_1, \ldots, p_{k+1}$ at pairwise distances $\Lambda$, and consider the following two neighboring datasets. The first dataset $S_1$ contains $n - k + 1$ copies of $p_1$, and (one copy of) $p_2, \ldots, p_k$. The second dataset $S_2$ is obtained from $S_1$ by replacing $p_k$ with $p_{k+1}$. Since in both cases there are only $k$ distinct input points, the optimal cost for each of these datasets is zero. On the other hand, By the constraint of differential privacy, the set of centers we compute essentially cannot be affected by this change. Therefore we expect that at least on one of these instances we will have error $\Omega(\Lambda^2)$. In what follows we will therefore assume that input points come from the $d$-dimensional ball $\mathcal{B}(0, \Lambda)$, and we assume that $\Lambda = 1$ in the introduction.

Traditionally, in the non-private literature, the goal is to minimize the multiplicative error $\gamma$, with the current state-of-the-art (non-private) algorithm achieving multiplicative error of $\gamma = 6.357$ (with no additive error) [2]. In contrast, in spite of the long line of works on private $k$-means [14, 38, 22, 25, 37, 45, 41, 43, 24, 8, 39, 29], all of the existing polynomial time private algorithms for the problem obtained only a super constant multiplicative error. We present the first polynomial time differentially private algorithm for the Euclidean $k$-means problem with constant multiplicative error, while essentially keeping the additive error the same as in previous state-of-the-art results. See Table 1 for a comparison.

## 1.1 Locally private $k$-means

In the local model of differential privacy (LDP), there are $n$ users and an untrusted server. Each user $i$ is holding a private input item $x_i$ (a point in $\mathbb{R}^d$ in our case), and the server's goal is to compute some function of the inputs (approximate the $k$-means in our case). However, in this model, the users do not send their data as is to the server. Instead, every user randomizes her data locally, and sends a differentially private report to the server, who aggregates all the reports. Informally, the privacy requirement is that the input of user $i$ has almost no effect on the distribution on the messages that user $i$ sends to the server. This is the model used by Apple, Google, and Microsoft in practice to ensure that private data never reaches their servers in the clear.

| Reference | Model | Multiplicative Error | Additive Error |
|---|---|---|---|
| Feldman et al. (2009) [22] | DP | $O(\sqrt{d})$ | $\tilde{O}\left((kd)^{2d}\right)$ |
| Nock et al. (2016) [41] | DP | $O(\log k)$ | $O\left(n/\log^2 n\right)$ |
| Feldman et al. (2017) [24] | DP | $O(k \log n)$ | $\tilde{O}\left(\sqrt{d} \cdot k^{1.5}\right)$ |
| Balcan et al. (2017) [8] | DP | $O(\log^3 n)$ | $\tilde{O}\left(d + k^2\right)$ |
| Nissim and Stemmer (2018) [39] | DP | $O(k)$ | $\tilde{O}\left(d^{0.51} \cdot k^{1.51}\right)$ |
| **This work** | **DP** | $\mathbf{O(1)}$ | $\tilde{\mathbf{O}}\left(\mathbf{k^{1.01} \cdot d^{0.51} + k^{1.5}}\right)$ |
| Nissim and Stemmer (2018) [39] | LDP $O(k \log n)$ rounds | $O(k)$ | $\tilde{O}\left(n^{0.67} \cdot d^{1/3} \cdot \sqrt{k}\right)$ |
| **This work** | **LDP** $\mathbf{O(1)}$ **rounds** | $\mathbf{O(1)}$ | $\tilde{\mathbf{O}}\left(\mathbf{n^{0.67} \cdot d^{1/3} \cdot k^2}\right)$ |
| Ahmadian et al. (2017) [2] | Non-private | 6.357 | 0 |

**Table 1**: Algorithms for $k$-means in the $d$-dimensional Euclidean space. *DP* denotes the standard (centralized) model of differential privacy, and *LDP* denotes the local model of differential privacy. Here $n$ is the number of input points, $k$ is the number of desired centers, and $d$ is the dimension. For simplicity, we assume that the input points come from the unit ball, and omit the dependency in $\varepsilon$, as well as logarithmic factors in $k, n, d, \beta, \delta$, from the additive error.

With increasing demand from the industry, the local model of differential privacy is now becoming more and more popular. Nevertheless, the only currently available $k$-means algorithm under LDP (with provable utility guarantees) is that of Nissim and Stemmer [39], with $O(k)$ multiplicative error. We present a new LDP algorithm for the $k$-means achieving *constant* multiplicative error. In addition, the protocol of [39] requires $O(k \log n)$ rounds of interaction between the server and the users, whereas our protocol uses only $O(1)$ rounds. See Table 1 for a comparison.

## 1.2 Classical algorithms are far from being private

We highlight some of the challenges that arise when trying to construct private variants for existing (non-private) algorithms. Recall for example the classical (non-private) Lloyd's algorithm, where in every iteration the input points are grouped by their proximity to the current centers, and the points in every group are averaged to obtain the centers for the next round. One barrier for constructing a private analogue for this algorithm is that, with differential privacy, the privacy parameters deteriorate with number of (private) computations that we apply to the dataset. So, even if we were able to construct a private analogue for every single iteration, our approximation guarantees would not necessarily improve with every iteration. In more details, composition theorems for differential privacy [21] allow for applying $O(n^2)$ private computations before exhausting the privacy budget completely. Lloyd's algorithm, however, might have a much larger number of iterations (exponential in $n$ in worst case). Even the bounds on it smoothed complexity are much larger than $n^2$ (currently $\approx n^{32}$ is known [5]). In addition, classical techniques for reducing the number of iterations often involve computations which are highly sensitive to a change of a small number of input points. For example, recall that in $k$-means++ [6] the initial $k$ centers (with which Lloyd's algorithm is typically initiated) are chosen *from the data points themselves*, an operation which cannot be applied as is when the data points are private.

These challenges are reflected in the recent work of Nock et al. [41], who constructed a private variant for the $k$-means++ algorithm. While their private algorithm achieves a relatively low multiplicative error of $O(\log k)$, their additive error is $\tilde{O}(n)$. In this work we are aiming for additive error at most polylogarithmic in $n$ (note that having additive error of $n$ is meaningless, since if points come from the unit ball then *every* choice of $k$ centers have error at most $O(n)$).

## 1.3 On the evolution of private $k$-means algorithms

The starting point of our work is the observation that by combining ideas from three previous works [25, 8, 39] we can obtain a differentially private $k$-means algorithm (in the centralized model) with constant multiplicative error, but with a relatively large additive error which is polynomial in $n$ (as we will see in Section 1.4). Most of our technical efforts (in the centralized model) are devoted to reducing the additive error while keeping the multiplicative error constant. We now describe the results of [25, 8, 39].

Gupta et al. [25] constructed a private variant for the classical local search heuristic [7, 32]. In this local search heuristic for $k$-means, we start with an arbitrary choice of $k$ centers, and then proceed in iterations, where in every iteration we replace one of our current centers with a new one, so as to reduce the $k$-means cost. Gupta et al. [25] constructed a private variant of the local search heuristic by instantiating the (generally inefficient) exponential mechanism of McSherry and Talwar [35] in order to privately choose a replacement center in every step. While the algorithm of Gupta et al. [25] obtains superb approximation guarantees, its runtime is exponential in the representation length of domain elements.[2] In particular, the algorithm is not applicable to the Euclidean space.

Balcan et al. [8] suggested the following strategy in order to adopt the techniques of Gupta et al. [25] to the Euclidean space. First, identify (in a differentially private manner) a small set $Y \subseteq \mathbb{R}^d$ of *candidate centers* such that $Y$ contains a subset of $k$ candidate centers with low $k$-means cost. Then, apply the techniques of Gupta et al. in order to choose $k$ centers from $Y$. If $|Y| = \text{poly}(n)$, then the resulting algorithm would be efficient. As the algorithm of Gupta et al. has very good approximation guarantees, the bottleneck for the approximation error in the algorithm of Balcan et al. is in the construction of $Y$. Namely, the overall error is dominated by the error of the best choice of $k$ centers out of $Y$ (compared to the cost of the best choice of $k$ centers from $\mathbb{R}^d$). Balcan et al. then constructed a differentially private algorithm for identifying a set of candidate centers $Y$ based on the Johnson-Lindenstrauss transform [31]. However, their construction gives a set of candidate centers such that the best choice of $k$ centers from these candidates is only guaranteed to have a multiplicative error of $O(\log^3 n)$, leading to a private $k$-means algorithm with $O(\log^3 n)$ multiplicative error.

---

[2]The algorithm of [25] obtains $O(1)$ multiplicative error and $\tilde{O}(k^2 d)$ additive error. It is designed for a *discrete* version of the problem, where input points come from a *finite* set $V$, and the running time of their algorithm is at least linear in $|V|$.

A different approach to obtain a good $k$-means clustering privately is via algorithms for the *1-cluster problem*, where given a set on $n$ input points in $\mathbb{R}^d$ and a parameter $t \leq n$, the goal is to identify a ball of the smallest radius that encloses at least $t$ of the input points. It was shown by Feldman et al. [24] that the Euclidean $k$-means problem can be reduced to the 1-cluster problem, by iterating the 1-cluster algorithm multiple times to find several balls that cover most of the data points. Feldman et al. then applied their reduction to the private 1-cluster algorithm of [40], and obtained a private $k$-means algorithm with multiplicative error $(k \log n)$. Following that work, Nissim and Stemmer [39] presented an improved algorithm for the 1-cluster problem which, when combined with the reduction of Feldman et al., gives a private $k$-means algorithm with multiplicative error $O(k)$.

## 1.4 Our techniques

Let $S \in (\mathbb{R}^d)^n$ be an input database and let $u_1^*, \ldots, u_k^* \in \mathbb{R}^d$ denote an optimal set of centers for $S$. We use $S_j^* \subseteq S$ to denote the cluster induced by $u_j^*$, i.e.,

$$S_j^* = \{x \in S : j = \mathrm{argmin}_\ell \|x - u_\ell^*\|\}.$$

We observe that the techniques that Nissim and Stemmer [39] applied to the 1-cluster problem can be used to privately identify a set of candidate centers $Y$ that "captures" every "big enough" cluster $j$. Informally, let $j$ be such that $|S_j^*| \geq n^a$ (for some constant $a > 0$). We will construct a set of candidate centers $Y$ such that there is a candidate center $y_j \in Y$ that is "close enough" to the optimal center $u_j^*$, in the sense that the cost of $y_j$ w.r.t. $S_j^*$ is at most a constant times bigger than the cost of $u_j^*$. That is, $\mathrm{cost}_{S_j^*}(\{y_j\}) = O\left(\mathrm{cost}_{S_j^*}(\{u_j^*\})\right)$. By simply ignoring clusters of smaller sizes, this means that $Y$ contains a subset $D$ of $k$ candidate centers such that

$$\mathrm{cost}_S(D) \leq O(1) \cdot \mathrm{OPT}_S + k \cdot n^a.$$

There are *two* reasons for the poly$(n)$ additive error incurred here. First, this technique effectively ignores every cluster of size less than $n^a$, and we pay $n^a$ additive error for every such cluster. Second, this technique only succeeds with polynomially small probability, and boosting the confidence using repetitions causes the privacy parameters to degrade.

We show that it is possible to boost the success probability of the above strategy without degrading the privacy parameters. To that end, we apply the repetitions to disjoint subsamples of the input points, and show that the subsampling process will not incur a poly$(n)$ error. In order to "capture" smaller clusters, we apply the above strategy repeatedly, where in every iteration we exclude from the computation the closest input points to the set of centers that we have already identified. We show that this technique allows to "capture" much smaller clusters. By combining this with the techniques of Balcan et al. and Gupta et al. for privately choosing $k$ centers out of $Y$, we get our new construction for $k$-means in the centralized model of differential privacy (see Table 1).

### 1.4.1 A construction for the local model

Recall that the algorithm of Gupta et al. (the private variant of the local search) applies the exponential mechanism of McSherry and Talwar [35] in order to privately choose a replacement center in every step. This use of the exponential mechanism is tailored to the centralized model, and it is not clear if the algorithm of Gupta et al. can be implemented in the local model. In addition, since the local search algorithm is iterative with a relatively large number of iterations (roughly $k \log n$ iterations), a local implementation of it, if exists, may have a large number of rounds of interaction between the users and the untrusted server.

To overcome these challenges, in our locally private algorithm for the $k$-means we first identify a set of candidate centers $Y$ (in a similar way to the centralized construction). Afterwards, we estimate the *weight* of every candidate center, where the *weight* of a candidate center $y$ is the number of input points $x \in S$ s.t. $y$ is the nearest candidate center to $x$. We show that the weighted set of candidate centers can be post-processed to obtain an approximation to the $k$-means of the input points. In order to estimate the weights we define a natural extension of the well-studied heavy-hitters problem under LDP, which reduces our incurred error. This results in our new construction for $k$-means in the local model of differential privacy (see Table 1).

### 1.4.2 Private coresets

A *coreset* [1] of set of input points $S$ is a small (weighted) set of points $P$ that captures some geometric properties of $S$. Coresets can be used to speed up computations, since if the coreset $P$ is much smaller than $S$, then optimization problems can be solved much faster by running algorithms on $P$ instead of $S$. In the context of $k$-means, the geometric property that we want $P$ to preserve is the $k$-means cost of *every* possible set of centers. That is, for every set of $k$ centers $D \subseteq \mathbb{R}^d$ we want that $\text{cost}_P(D) \approx \text{cost}_S(D)$ (where in $\text{cost}_P(D)$ we multiply each distance by the weight of the corresponding point). Coresets for $k$-means and $k$-medians have been the subject of many recent papers, such as [17, 23, 26, 27, 10, 18]. *Private* coresets for $k$-means and $k$-medians have been considered in [22] and in [24]. We show that our techniques result in new constructions for private coresets for $k$-means and $k$-medians, both for the centralized and for the local model of differential privacy. In the local model, this results in the *first* private coreset scheme with provable utility guarantees. In the centralized model, our new construction achieves significantly improved error rates over the previous state-of-the-art.

## 2 Preliminaries

In $k$-means clustering we aim to partition $n$ points into $k$ clusters in which each point $x$ belongs to the cluster whose mean is closest to $x$. Formally, for a set of points $S \in (\mathbb{R}^d)^n$ and a set of centers $C \subseteq \mathbb{R}^d$, the *cost* of $C$ w.r.t. the points $S$ is defined as

$$\text{cost}_S(C) = \sum_{x \in S} \min_{c \in C} \|x - c\|_2^2.$$

For a *weighted* set $S = \{(x_1, \alpha_1), \ldots, (x_n, \alpha_n)\} \in (\mathbb{R}^d \times \mathbb{R})^n$, the *weighted cost* is

$$\text{cost}_S(C) = \sum_{(x, \alpha) \in S} \alpha \cdot \min_{c \in C} \|x - c\|_2^2.$$

**Definition 2.1** (*k*-means). *Let $S$ be a (weighted or unweighted) finite set of points in $\mathbb{R}^d$. A set $C^*$ of $k$ centers in $\mathbb{R}^d$ is called k-means of $S$ if it minimizes $\text{cost}_S(C)$ over every such set $C$.*

For a set of points $S \in (\mathbb{R}^d)^n$ we use $\text{OPT}_S$ to denote the cost of the *k*-means of $S$. That is,

$$\text{OPT}_S = \min_{\substack{C \subseteq \mathbb{R}^d \\ |C|=k}} \{\text{cost}_S(C)\}.$$

Moreover, for a set of points $S \in (\mathbb{R}^d)^n$ and a set of centers $Y \subseteq \mathbb{R}^d$ we write $\text{OPT}_S(Y)$ to denote the lowest possible cost of $k$ centers from $Y$. That is,

$$\text{OPT}_S(Y) = \min_{\substack{C \subseteq Y \\ |C|=k}} \{\text{cost}_S(C)\}.$$

**Definition 2.2** (Approximated *k*-means). *Let $S$ be a (weighted or unweighted) finite set of points in $\mathbb{R}^d$. A set $C$ of $k$ centers in $\mathbb{R}^d$ is a $(\gamma, \eta)$-approximation for the k-means of $S$ if*

$$\text{cost}_S(C) \leq \gamma \cdot \text{OPT}_S + \eta.$$

## 2.1 Preliminaries from differential privacy

Consider a database where each entry contains information pertaining to one individual. An algorithm operating on databases is said to preserve differential privacy if a change of a single record of the database does not significantly change the output distribution of the algorithm. Intuitively, this means that individual information is protected: whatever is learned about an individual could also be learned with her data arbitrarily modified (or without her data at all).

**Definition 2.3** (Differential Privacy [20]). *Two databases $S, S' \in X^n$ are called* neighboring *if they differ in at most one entry. A randomized algorithm $M : X^n \to Y$ is $(\varepsilon, \delta)$ differentially private if for every two neighboring datasets $S, S' \in X^n$ and every $T \subseteq Y$ we have*

$$\Pr[M(S) \in T] \leq e^\varepsilon \cdot \Pr[M(S') \in T] + \delta,$$

*where the probability is over the randomness of $M$.*

### 2.1.1 The Laplace and Gaussian mechanisms

The most basic constructions of differentially private algorithms are via the Laplace and Gaussian mechanisms as specified in the following theorems.

**Definition 2.4** ($L_p$-Sensitivity). *A function $f$ mapping databases to $\mathbb{R}^d$ has $L_p$-sensitivity $\lambda$ if $\|f(S) - f(S')\|_p \leq \lambda$ for all neighboring $S, S'$.*

**Theorem 2.5** (Laplace mechanism [20]). *A random variable is distributed as $\text{Lap}(b)$ if its probability density function is $h(y) = \frac{1}{2b}\exp(-\frac{|y|}{b})$. Let $\varepsilon > 0$, and let $f : U^n \to \mathbb{R}^d$ be a function of $L_1$-sensitivity $\lambda$. The mechanism $\mathcal{A}$ that on input $S \in U^n$ outputs $f(S) + \left(\text{Lap}(\frac{\lambda}{\varepsilon})\right)^d$ is $(\varepsilon, 0)$-differentially private. Moreover,*

$$\Pr\left[\|\mathcal{A}(S) - f(S)\|_\infty > \Delta\right] \leq d \cdot \exp\left(-\frac{\varepsilon \Delta}{\lambda}\right).$$

**Theorem 2.6** (Gaussian Mechanism [19]). *Let $\varepsilon, \delta \in (0,1)$, and assume $f : U^n \to \mathbb{R}^d$ has $L_2$-sensitivity $\lambda$. Let $\sigma \geq \frac{\lambda}{\varepsilon}\sqrt{2\ln(1.25/\delta)}$. The mechanism that on input $S \in U^n$ outputs $f(S) + \left(\mathcal{N}(0, \sigma^2)\right)^d$ is $(\varepsilon, \delta)$-differentially private.*

### 2.1.2 Noisy average of vectors in $\mathbb{R}^d$

Consider the task of privately estimating the average of $n$ vectors in the $d$-dimensional ball $\mathcal{B}(0, \Lambda)$. The Gaussian mechanism (Theorem 2.6) allows for privately estimating this average with additive $L_2$ error $\approx \frac{\Lambda \cdot \sqrt{d}}{\varepsilon n}$. In some cases, we could relax the dependency on $\Lambda$ using the following theorem.

**Theorem 2.7** ([40]). *Let $\beta, \varepsilon, \delta > 0$. There exists an efficient $(\varepsilon, \delta)$-differentially private algorithm that takes a database $S \in (\mathbb{R}^d)^n$ and a parameter $r$. The algorithm outputs a point $y \in \mathbb{R}^d$ such that if $\mathrm{diam}(S) \leq r$ then with probability at least $(1 - \beta)$ it holds that*

$$\left\| y - \frac{1}{n} \sum_{x \in S} x \right\|_\infty \leq O\left( \frac{r}{\varepsilon n} \ln\left(\frac{nd}{\beta}\right) \sqrt{\ln\left(\frac{1}{\delta}\right)} \right).$$

*Importantly, differential privacy is guaranteed to hold even if $\mathrm{diam}(S) > r$.*

Let $0 < \alpha < 1$ be a parameter, and observe that if $n \gtrsim \frac{\sqrt{d}}{\alpha \varepsilon}$ then the algorithm from the theorem above returns (w.h.p.) an estimation for the average of $S$ with $L_2$ error at most $\alpha r$.

### 2.1.3 Private $k$-means and candidate centers

In the discrete version of the $k$-means clustering problem, there is a fixed and finite subset $Y \subseteq \mathbb{R}^d$, which we call *candidate centers*. Given a set of points $S \in (\mathbb{R}^d)^n$ our goal is to identify a subset $C \subseteq Y$ of size $k$ with the lowest possible cost. That is, instead of searching for $k$ centers in $\mathbb{R}^d$, we are searching for $k$ centers in $Y$, and our runtime is allowed to depend polynomially on $|Y|$. As was shown by Gupta et al. [25] and Balcan et al. [8], it is possible to *privately* approximate this discrete version of the problem by constructing a differentially private variant of the local search algorithm [32].

**Theorem 2.8** ([25, 8]). *Let $\beta, \varepsilon, \delta > 0$ and $k \in \mathbb{N}$, and let $Y \subseteq \mathbb{R}^d$ be a finite set centers. There exists an $(\varepsilon, \delta)$-differentially private algorithm that takes a database $S$ containing $n$ points from the $d$-dimensional ball $\mathcal{B}(0, \Lambda)$, and outputs a subset $D \subseteq Y$ of size $|D| = k$ s.t. with probability at least $(1 - \beta)$ we have that*

$$\mathrm{cost}_S(D) \leq O(1) \cdot \mathrm{OPT}_S(Y) + O\left( \frac{k^{1.5} \Lambda^2}{\varepsilon} \log\left(\frac{n|Y|}{\beta}\right) \sqrt{\log(n) \cdot \log\left(\frac{1}{\delta}\right)} \right).$$

In [25, 8] the above theorem is stated slightly differently, for pure $(\varepsilon, 0)$-differential privacy. The variant stated in Theorem 2.8 results from the stronger composition properties of $(\varepsilon, \delta)$-differential privacy (see [21]).

## 2.2 Locality sensitive hashing

A locality sensitive hash function aims to maximize the probability of a collision for similar items, while minimizing the probability of collision for dissimilar items. Formally,

**Definition 2.9** ([30]). *Let $\mathcal{M}$ be a metric space, and let $r > 0$, $c > 1$, $0 \leq q < p \leq 1$. A family $\mathcal{H}$ of functions mapping $\mathcal{M}$ into domain $U$ is an $(r, cr, p, q)$ locality sensitive hashing family (LSH) if for all $x, y \in \mathcal{M}$ (i) $\Pr_{h \in_R \mathcal{H}}[h(x) = h(y)] \geq p$ if $d_\mathcal{M}(x, y) \leq r$; and (ii) $\Pr_{h \in_R \mathcal{H}}[h(x) = h(y)] \leq q$ if $d_\mathcal{M}(x, y) \geq cr$.*

## 2.3 The Poisson approximation

When throwing $n$ balls into $R$ bins, the distribution of the number of balls in a given bin is $\text{Bin}(n, 1/R)$. As the Poisson distribution is the limit distribution of the binomial distribution when $n/R$ is fixed and $n \to \infty$, the distribution of the number of balls in a given bin is approximately $\text{Pois}(n/R)$. In fact, in some cases we may approximate the *joint distribution* of the number of balls in all the bins by assuming that the load in each bin is an independent Poisson random variable with mean $n/R$.

**Theorem 2.10** (e.g., [36]). *Suppose that $n$ balls are thrown into $R$ bins independently and uniformly at random, and let $X_i$ be the number of balls in the $i^{th}$ bin, where $1 \leq i \leq R$. Let $Y_1, \cdots, Y_R$ be independent Poisson random variables with mean $n/R$. Let $f(x_1, \cdots, x_R)$ be a non-negative function. Then,*
$$\mathbb{E}\left[f(X_1, \cdots, X_R)\right] \leq e\sqrt{n}\, \mathbb{E}\left[f(Y_1, \cdots, Y_R)\right].$$

In particular, the theorem states that any event that takes place with probability $p$ in the Poisson case, takes place with probability at most $pe\sqrt{n}$ in the exact case (this follows by letting $f$ be the indicator function of that event).

We will also use the following bounds on the tail probabilities of a Poisson random variable:

**Theorem 2.11** ([3]). *Let $X$ have Poisson distribution with mean $\mu$. For $0 \leq \alpha \leq 1$,*
$$\begin{aligned}
\Pr[X \leq \mu(1-\alpha)] &\leq e^{-\alpha^2 \mu/2} \\
\Pr[X \geq \mu(1+\alpha)] &\leq e^{-\alpha^2 \mu/3}.
\end{aligned}$$

# 3 Private $k$-means – the centralized setting

In this section we present our construction for the centralized model. The main step in the construction is to identify a set of candidate centers that contains a subset of $k$ centers with low $k$-means cost. Consider an input database $S$, and let $u_1^*, \ldots, u_k^* \in \mathbb{R}^d$ denote an optimal set of $k$ centers for $S$. In Section 3.1 we use the techniques of Nissim and Stemmer [39] in order to identify a set of candidate centers that contains a "close enough" candidate center to every optimal center $u_j^*$, provided that the optimal cluster induced by $u_j^*$ is "big enough". Smaller clusters will be handled in Section 3.2.

## 3.1 Reformulating the results of [39]

In this section we present a procedure, named `LSH-Procedure`, for privately identifying a set of centers with a small distance to *some* of the input points. Most of the ideas in the analysis of this procedure have appeared in the work of Nissim and Stemmer [39] who studied the related *1-Cluster* problem. We modify their procedure to output a set of several *candidate centers*, and boost the success probability. We will later apply this procedure iteratively in our construction for approximating the $k$-means.

We use a family $\mathcal{H}$ of $(r, cr, p{=}n^{-b}, q{=}n^{-2-a})$-locality sensitive hash functions, mapping $\mathbb{R}^d$ to a universe $U$, for some constants $1 > a > b > 0$, $r > 0$, and $c > 1$. Such families exist for every choice of constants $1 > a > b > 0$ and $r > 0$, with $c = c(a, b)$ (see, e.g., [4]). Furthermore, w.l.o.g., we can assume that the range $U$ of the functions in $\mathcal{H}$ is of size $|U| \leq n^3$. If this is not the case, then we

**Algorithm LSH-Procedure**

**Input:** Database $S \in \mathbb{R}^d$ containing $n$ points, number $r > 0$, failure probability $\beta$, privacy parameters $\varepsilon, \delta$.

**Tool used:** Family $\mathcal{H}$ of $(r, cr, p=n^{-b}, q=n^{-2-a})$-locality sensitive hash functions mapping $\mathbb{R}^d$ to a universe $U$, for some parameters $0 < b < a < 1$, and $c = c(a, b) > 1$.

1. Denote $M = 2n^a \ln(\frac{1}{\beta})$, and randomly partition $S$ into $S_1, S_2, \ldots, S_M$.

2. Sample $M$ hash function $h_1, \ldots, h_M \in \mathcal{H}$ mapping $\mathbb{R}^d$ to $U$. For $(m, u) \in [M] \times U$ define $S_{m,u}$ as the multiset containing all elements of $S_m$ that are mapped into $u$ by $h_m$, i.e.,
$$S_{m,u} \triangleq \{x \in S_m \ : \ h_m(x) = u\}.$$

3. Use the Laplace mechanism[3] with privacy parameter $\frac{\varepsilon}{2}$ to obtain estimations $\hat{w}_{m,u} \approx |S_{m,u}|$ for every $(m, u) \in [M] \times U$. Denote $L = \left\{(m, u) : \hat{w}_{m,u} \geq \frac{60}{\varepsilon} \ln(\frac{n}{\beta})\right\}$.

4. For every $(m, u) \in L$, use the algorithm from Theorem 2.7 on the database $S_{m,u}$ with privacy parameters $(\frac{\varepsilon}{2}, \delta)$ and $cr$ as a bound on $\text{diam}(S_{m,u})$ to obtain a point $\hat{y}_{m,u}$ approximating the average of $S_{m,u}$.

5. Output $\{\hat{y}_{m,u} : (m, u) \in L\}$

---

can simply apply a (pairwise independent) hash function with range $n^3$ to the output of the locally sensitive hash function. Clearly, this does not decrease the collision probability of "close" elements (within distance $r$), and moreover, this can increase the collision probability of "non-close" elements (at distance at least $cr$) by at most $n^{-3} = o(n^{-2-a}) = o(q)$.

**Lemma 3.1.** *Algorithm LSH-Procedure is $(\varepsilon, \delta)$-differentially private. Furthermore, there exists a constant $\Gamma > 1$ such that the following holds. Assume we apply LSH-Procedure to a database $S \in (\mathbb{R}^d)^n$ with parameters $r, \beta, \varepsilon, \delta, a, b$, and $c$. Let $P \subseteq S$ be s.t. $\text{diam}(P) \leq r$ and $|P| = t$ for some $t$ satisfying*
$$t \geq \frac{\Gamma}{\varepsilon} \cdot \sqrt{d} \cdot n^{a+b} \cdot \ln\left(\frac{1}{\beta}\right) \ln\left(\frac{n}{\beta}\right) \sqrt{\ln\left(\frac{1}{\delta}\right)}.$$

*The algorithm outputs a set of at most $\varepsilon n/\log n$ centers, s.t. with probability at least $1 - \beta$ a ball of radius $(2c + 1)r$ around one of these centers contains all of $P$.*

The proof is very similar to the that of [39]. We include the proof in the appendix for completeness.

**Remark 3.2.** *We think of $a$ and $b$ as small constants, e.g., $a = 0.2$ and $b = 0.1$. Hence, ignoring logarithmic factors, the above theorem only requires $t$ to be as big as $n^{0.3}\sqrt{d}/\epsilon$. Smaller constants $a, b$ result in a bigger (but constant) approximation factor $c = c(a, b)$.*

---

[3] As described, this step requires runtime of roughly $M \cdot |U| = \text{poly}(n)$. We remark that it is possible to reduce the runtime to almost linear in $n$ by using more efficient algorithms for computing histograms. See e.g., [13, 16, 44, 9].

---

**Algorithm `Private-Centers`**

**Input:** Database $S$ containing $n$ points in the $d$-dimensional ball $\mathcal{B}(0,\Lambda)$, failure probability $\beta$, privacy parameters $\varepsilon, \delta$, and additional LSH parameters $0 < b < a < 1$, and $c = c(a,b) > 1$ (as in Algorithm `LSH-Procedure`).

1. Initiate $C = \emptyset$.

2. For $r = \frac{\Lambda}{n}, \frac{2 \cdot \Lambda}{n}, \ldots, \frac{2^i \cdot \Lambda}{n}, \ldots, \Lambda$:
   Run algorithm `LSH-Procedure` on the database $S$ with parameters $\frac{\varepsilon}{\log n}, \frac{\delta}{\log n}, r, \beta, a, b, c$, and add the returned set of centers to $C$.

3. Output $C$.

---

Recall that algorithm `LSH-Procedure` requires an input parameter $r$ that bounds the diameter of a subset of $t$ input points. The next construction removes the necessity of this input by executing the `LSH-Procedure` multiple times with exponentially growing choices for the parameter $r$.

**Lemma 3.3.** *Algorithm `Private-Centers` is $(\varepsilon, \delta)$-differentially private. Furthermore, there exists a constant $\Gamma > 1$ such that the following holds. Assume we apply `Private-Centers` to a database $S$ containing $n$ points in the $d$-dimensional ball $\mathcal{B}(0,\Lambda)$, with parameters $\beta, \varepsilon, \delta, a, b, c$. Let $P \subseteq S$ be s.t. $|P| = t$ for some $t$ satisfying*

$$t \geq \frac{\Gamma}{\varepsilon} \cdot \sqrt{d} \cdot n^{a+b} \cdot \log(n) \cdot \ln\left(\frac{1}{\beta}\right) \ln\left(\frac{n}{\beta}\right) \sqrt{\ln\left(\frac{\log n}{\delta}\right)}.$$

*The algorithm outputs a set of at most $\varepsilon n$ centers, s.t. with probability at least $1 - \beta$ a ball of radius $O(\operatorname{diam}(P) + \frac{\Lambda}{n})$ around one of these centers contains all of $P$.*

*Proof.* The privacy guarantees of the algorithm are immediate. As for the utility analysis, recall that in Step 2 the algorithm applies the `LSH-Procedure` with exponentially growing choices of $r$. Let $i^*$ be the smallest integer s.t. $r^* = \frac{2^{i^*} \cdot \Lambda}{n} \geq \operatorname{diam}(P)$. Clearly, $r^* \leq 2 \cdot \operatorname{diam}(P) + \frac{\Lambda}{n}$. Now consider the application of `LSH-Procedure` in Step 2 of algorithm `Private-Centers`, in which $r = r^*$. The statement now follows from Lemma 3.1. □

**Remark 3.4.** *Using standard techniques for confidence amplification (applying `Private-Centers` with a constant confidence parameter $\log(1/\beta)$ times), the requirement on $t$ in Lemma 3.3 can be replaced with*

$$t \geq \frac{\Gamma}{\varepsilon} \cdot \sqrt{d} \cdot n^{a+b} \cdot \ln\left(\frac{1}{\beta}\right) \ln^2(n) \sqrt{\ln\left(\frac{\log n}{\delta}\right)},$$

*at the expense of returning $\varepsilon n \log(1/\beta)$ centers instead of $\varepsilon n$ centers. Furthermore, by slightly increasing the constant $a$, the requirement on $t$ can be written as*

$$t \geq \frac{\Gamma}{\varepsilon} \cdot \sqrt{d} \cdot n^{a+b} \cdot \ln\left(\frac{1}{\beta}\right) \sqrt{\ln\left(\frac{1}{\delta}\right)}.$$

**Algorithm** `Private-`$k$`-Means`

**Input:** Database $S$ containing $n$ points in the $d$-dimensional ball $\mathcal{B}(0,\Lambda)$, failure probability $\beta$, privacy parameters $\varepsilon, \delta$, and additional LSH parameters $0 < b < a < 1$, and $c = c(a,b) > 1$ (as in Algorithm `Private-centers`).

% Let $u_1^*, \ldots, u_k^*$ denote an optimal set of centers for $S$, and let $S_j^*$ be the cluster induced by $u_j^*$, i.e., $S_j^* = \{x \in S : j = \operatorname{argmin}_\ell \|x - u_\ell^*\|\}$. For $j \in [k]$ let $r_j^* = \sqrt{\frac{2}{|S_j^*|} \sum_{x \in S_j^*} \|x - u_j^*\|^2}$, and let $P_j^* = \mathcal{B}(u_j^*, r_j^*) \cap S_j^*$.

1. Initiate $C = \emptyset$, and denote $S_1 = S$ and $n_1 = n$.

% Initiate `ASSIGN`$[j] = \bot$ for every $j \in [k]$.

2. For $i = 1$ to $\log \log n$ do

   (a) Run algorithm `Private-Centers` on the database $S_i$ with parameters $\frac{\varepsilon}{\log \log n}, \frac{\delta}{\log \log n}, \frac{\beta}{k}, a, b, c$, and add the returned set of centers to $C$.

   % For every $j \in [k]$: if `ASSIGN`$[j] = \bot$ and if $\exists u_j \in C$ s.t. $\|u_j - u_j^*\| \leq O(r_j^* + \frac{\Lambda}{n})$, then set `ASSIGN`$[j] = u_j$.

   (b) Let $S_{i+1} \subseteq S_i$ be a subset of $S_i$ containing $n_{i+1} = 2(T+1)wk \cdot n_i^{a+b}$ points with the largest distance to the centers in $C$, where $w = w(n, d, k, \beta, \varepsilon, \delta)$ and $T = T(n)$ will be specified in the analysis.

   % For every $j \in [k]$: if `ASSIGN`$[j] = \bot$ and if $P_j^* \not\subseteq S_{i+1}$, then let $p_j \in P_j^* \setminus S_{i+1}$, let $u_j = \operatorname{argmin}_{u \in C} \|p_j - u\|$, and set `ASSIGN`$[j] = u_j$.

3. Output $C$.

% For every $j \in [k]$: if `ASSIGN`$[j] = \bot$, then arbitrarily choose $u_j \in C$ and set `ASSIGN`$[j] = u_j$.

## 3.2 Capturing smaller and smaller clusters

We are now ready to present our construction for the centralized model – algorithm `Private-`$k$`-Means`. The algorithm privately identifies a polynomial set of candidate centers such that there exists a subset of $k$ candidate centers with low $k$-means cost. By the results of Gupta et al. [25] and Balcan et al. [8], this suffices for privately approximating the $k$-means of the inputs (see Theorem 2.8).

For readability, we have added inline comments throughout the description of `Private-`$k$`-Means`, which will be helpful for the analysis. These comments are not part of the algorithm. Let $u_1^*, \ldots, u_k^*$ denote an optimal set of centers w.r.t. the set of input points $S$, and let $S_1^*, \ldots, S_k^* \subseteq S$ be the clusters induced by these optimal centers, i.e., $S_j^* \subseteq S$ is the set of input points whose nearest optimal center is $u_j^*$. (These optimal centers and clusters are *unknown* to the algorithm; they are only used in the inline comments and in the analysis.) Throughout the execution, we use the inline comments in order to prescribe a feasible (but not necessarily optimal) assignment of the data points to (a subset of $k$ of) the current candidate centers. Specifically, we maintain an array `ASSIGN`, where we write `ASSIGN`$[j] = u$ (for some center $u$ in our current set of candidate centers) to denote that *all* of the points in the optimal cluster $S_j^*$ are assigned to the candidate center $u$. We write `ASSIGN`$[j] = \bot$ to denote that points in $S_j^*$ have not been assigned to a center yet. For every $j$ we have that `ASSIGN`$[j] = \bot$ at the beginning of the execution, and that `ASSIGN`$[j]$ is changed exactly once during the execution, at which point the $j$th cluster is *assigned* to a center. In the analysis we argue that at the end of the execution the resulting assignment has low $k$-means cost.

| | |
|---|---|
| $S$ | The input database. |
| $u_1^*, \ldots, u_k^* \in \mathbb{R}^d$ | An optimal set of centers for $S$. |
| $S_1^*, \ldots, S_k^* \subseteq S$ | The clusters induced by $u_1^*, \ldots, u_k^*$. |
| $r_1^*, \ldots, r_k^* \in \mathbb{R}^{\geq 0}$ | $r_j^* = \sqrt{\frac{2}{|S_j^*|} \sum_{x \in S_j^*} \|x - u_j^*\|^2}$. |
| $P_1^*, \ldots, P_k^*$ | $P_j^* = \mathcal{B}(u_j^*, r_j^*) \cap S_j^*$. |
| $S_i \subseteq S$, $i \in [\log \log n]$ | The set of remaining input points during the $i$th iteration. |
| $n_i = |S_i|$, $i \in [\log \log n]$ | The number of remaining input points during the $i$th iteration. |
| $C$ | The current set of candidate centers. |
| $\text{ASSIGN}[j]$, $j \in [k]$ | The assignment constructed in the inline comments. |

**Table 2**: Notations for the analysis of algorithm Private-$k$-Means

**Notation.** For a point $x \in S$, we write $\text{ASSIGN}(x)$ to denote the candidate center to which $x$ is assigned at a given moment of the execution. That is, $\text{ASSIGN}(x) = \text{ASSIGN}[j]$, where $j$ is s.t. $x \in S_j^*$.

Consider the execution of algorithm Private-$k$-Means. For readability, we have summarized some of the notations that are specified in the algorithm in Table 2. In addition, we denote

$$w = \frac{\Gamma \cdot \sqrt{d}}{\varepsilon} \cdot \log \log(n) \cdot \log\left(\frac{k}{\beta}\right) \sqrt{\log\left(\frac{\log \log n}{\delta}\right)}, \tag{1}$$

where $\Gamma$ is the constant from Remark 3.4. Consider the following good event:

> **Event $E_1$ (over the randomness of Private-Centers):**
> For every $i \in [\log \log n]$ and $j \in [k]$, if $|P_j^* \cap S_i| \geq w \cdot n_i^{a+b}$ then after Step 2a of the $i$th iteration, the set $C$ contains a center $u_j \in C$ s.t. $\|u_j - u_j^*\| \leq O(r_j^* + \frac{\Lambda}{n})$.

**Claim 3.5.** *Event $E_1$ occurs with probability at least $1 - \beta$.*

*Proof.* By the properties of algorithm Private-Centers (Lemma 3.3), and a union bound over $i \in [\log \log n]$ and $j \in [k]$, Event $E_1$ happens with probability at least $1 - \beta$. □

The proof continues by showing that whenever event $E_1$ occurs, algorithm Private-$k$-Means successfully identifies a "good" set of candidate centers, in the sense that there is a subset of $k$ candidate centers with low $k$-means cost w.r.t. the input $S$. We first show that if $E_1$ occurs then the number of unassigned points reduces quickly in every iteration.

**Claim 3.6.** *If Event $E_1$ occurs, then for every $i \in [\log \log n]$, before Step 2b of the $i$th iteration there are at most $2kw \cdot n_i^{a+b}$ unassigned points in $S$, i.e., $|\{x \in S : \text{ASSIGN}(x) = \bot\}| \leq 2kw \cdot n_i^{a+b}$.*

*Proof.* Observe that every cluster $\ell$ with $|P_\ell^*| \geq w \cdot n_i^{a+b}$ is assigned before Step 2b of the $i$th iteration, because if $P_\ell^* \subseteq S_i$ then by Event $E_1$ cluster $\ell$ is assigned after Step 2a of the $i$th iteration, and if $P_\ell^* \not\subseteq S_i$ then the assignment must have already occurred before the $i$th iteration. Recall that $|P_\ell^*| \geq \frac{1}{2}|S_\ell^*|$, as otherwise less than half of the points in $S_\ell^*$ are within distance $r_\ell^*$ from $u_\ell^*$, and so $\text{cost}_{S_\ell^*}(\{u_\ell^*\}) > \frac{|S_\ell^*|}{2} \cdot (r_\ell^*)^2 = \text{cost}_{S_\ell^*}(\{u_\ell^*\})$. Hence, before Step 2b of the $i$th iteration, there could be at most $2kw \cdot n_i^{a+b}$ unassigned points. □

**Notation.** For $i \in [\log \log n]$ we denote by $A_i \subseteq S$ and $B_i \subseteq S$ the subset of input points whose cluster is assigned to a center during the $i$th iteration in the comments after Step 2a and after Step 2b, respectively. Observe that $A_1, B_1, \ldots, A_{\log \log n}, B_{\log \log n}$ are mutually disjoint.

**Notation.** Recall the optimal centers $u_1^*, \ldots, u_k^*$ and the radiuses $r_1^*, \ldots, r_k^*$ defined in the first comment in algorithm Private-$k$-Means. For a point $x \in \mathbb{R}^d$, let $u^*(x)$ denote $x$'s nearest optimal center, and $r^*(x)$ its corresponding radius.

**Observation 3.7.** *For every $i \in [\log \log n]$ and for every $x \in A_i$, at the end of the execution we have*

$$\|x - \mathtt{ASSIGN}(x)\|^2 \leq O\left(\|x - u^*(x)\|^2 + (r^*(x))^2 + \frac{\Lambda^2}{n^2}\right).$$

**Lemma 3.8.** *If Event $E_1$ occurs, then for every iteration $i \in [\log \log n]$ and for every $x \in B_i$ there exists a set of input points $Q(x) \subseteq S$ such that*

1. *For every $i \in [\log \log n]$ and for every $x \in B_i$ it holds that $|Q(x)| = T$, where $T = O(\log \log n)$.*

2. *For every $i \in [\log \log n]$ and for every $x, y \in B_i$, if $x \neq y$ then $Q(x) \cap Q(y) = \emptyset$.*

3. *For every $i \in [\log \log n]$ and for every $x \in B_i$, at the end of the execution it holds that*

$$\|x - \mathtt{ASSIGN}(x)\|^2 \leq O\left(\|x - u^*(x)\|^2 + (r^*(x))^2 + \frac{1}{T}\sum_{q \in Q(x)} \|q - \mathtt{ASSIGN}(q)\|^2\right).$$

*Proof.* Let us focus on the $i$th iteration. By Claim 3.6, before Step 2b, there could be at most $2k \cdot w \cdot n_i^{a+b}$ unassigned points in $S$. In particular, $|B_i| \leq 2k \cdot w \cdot n_i^{a+b}$. As $|S_{i+1}| = n_{i+1} = 2(T+1)wk \cdot n_i^{a+b}$, we have that $S_{i+1}$ contains at least $2Twk \cdot n_i^{a+b}$ assigned points (which were already assigned before the $i$th iteration). For $x \in B_i$ we define $Q(x)$ to be an arbitrary set of $T$ assigned points from $S_{i+1}$, such that for all $x \neq y \in B_i$ it holds that $Q(x) \cap Q(y) = \emptyset$. It remains to prove item 3 of the lemma.

**Notation.** We write $Q_i$ to denote the union of all sets $Q(x)$ defined during the $i$th iteration. That is, $Q_i = \bigcup_{x \in B_i} Q(x)$.

Let $j$ be a cluster that is assigned to a center during the $i$th iteration in the comment following Step 2b (so $S_j^* \subseteq B_i$). As in the comment, let $p_j \in P_j^* \setminus S_{i+1}$, and $u_j = \operatorname{argmin}_{u \in C} \|p_j - u\|$, where $C$ is the set of candidate centers at that time. Recall that we set $\mathtt{ASSIGN}[j] = u_j$, and observe that since $p_j \notin S_{i+1}$, for every $q \in Q_i \subseteq S_{i+1}$, we have that $\|p_j - u_j\|^2 \leq \min_{c \in C} \|q - c\|^2$. In particular, since every $q \in Q_i$ was already assigned to a center before the $i$th iteration, we have that $\mathtt{ASSIGN}(q) \in C$, and hence $\|p_j - u_j\|^2 \leq \|q - \mathtt{ASSIGN}(q)\|^2$.

Now let $x \in S_j^*$. We get that

$$\|x - \texttt{ASSIGN}(x)\|^2 = \|x - u_j\|^2 \leq O\left(\|x - u_j^*\|^2 + \|u_j^* - p_j\|^2 + \|p_j - u_j\|^2\right)$$

$$\leq O\left(\|x - u_j^*\|^2 + \|u_j^* - p_j\|^2 + \frac{1}{T}\sum_{q \in Q(x)} \|q - \texttt{ASSIGN}(q)\|^2\right)$$

$$\leq O\left(\|x - u_j^*\|^2 + (r_j^*)^2 + \frac{1}{T}\sum_{q \in Q(x)} \|q - \texttt{ASSIGN}(q)\|^2\right).$$

□

**Lemma 3.9.** *Algorithm* `Private-k-Means` *is* $(\varepsilon, \delta)$*-differentially private. Furthermore, if the algorithm is applied to a database $S$ containing $n$ points in the $d$-dimensional ball $\mathcal{B}(0, \Lambda)$, then it outputs a set of at most $\varepsilon n \log(\frac{k}{\beta})$ centers, s.t. with probability at least $1 - \beta$*

$$\min_{\substack{D \subseteq C \\ |D|=k}} \{\text{cost}_S(D)\} \leq O(1) \cdot \text{OPT}_S + O\left(T^{\frac{1}{1-a-b}} \cdot w^{\frac{1}{1-a-b}} \cdot k^{\frac{1}{1-a-b}}\right) \cdot \Lambda^2,$$

*where $w$ is defined in Equation (1), and $T = \Theta(\log \log n)$.*

*Proof.* We show that the stated bound holds for the assignment described in the inline comments throughout the algorithm (the array `ASSIGN`) at the end of the execution. First observe that by Claim 3.6 and by the fact that there are $\log \log n$ iterations, at the end of the execution there could be at most $O\left((2(T+1)wk)^{\frac{1}{1-a-b}}\right)$ unassigned input points. Let us denote the set of unassigned points as $H$. The distance from each unassigned point to an arbitrary center is trivially at most $\Lambda$. For every assigned point $x$, by Observation 3.7 and by Lemma 3.8, either $\|x - \texttt{ASSIGN}(x)\|^2 = O(\|x - u^*(x)\|^2 + (r^*(x))^2 + \frac{\Lambda^2}{n^2})$, or

$$\|x - \texttt{ASSIGN}(x)\|^2 \leq O\left(\|x - u^*(x)\|^2 + (r^*(x))^2 + \frac{1}{T}\sum_{q \in Q(x)} \|q - \texttt{ASSIGN}(q)\|^2\right).$$

Hence,

$$\text{cost}_S(\{\texttt{ASSIGN[j]} : j \in [k]\}) = \sum_{x \in S} \|x - \texttt{ASSIGN}(x)\|^2$$

$$= \sum_{x \in H} \|x - \texttt{ASSIGN}(x)\|^2 + \sum_{\substack{i \in [\log \log n] \\ x \in A_i}} \|x - \texttt{ASSIGN}(x)\|^2 + \sum_{\substack{i \in [\log \log n] \\ x \in B_i}} \|x - \texttt{ASSIGN}(x)\|^2$$

$$\leq O\left((2(T+1)wk)^{\frac{1}{1-a-b}}\right) \cdot \Lambda^2 + \sum_{\substack{i \in [\log \log n] \\ x \in A_i}} O\left(\|x - u^*(x)\|^2 + (r^*(x))^2 + \frac{\Lambda^2}{n^2}\right)$$

$$+ \sum_{\substack{i \in [\log \log n] \\ x \in B_i}} O\left(\|x - u^*(x)\|^2 + (r^*(x))^2 + \frac{1}{T}\sum_{q \in Q(x)} \|q - \texttt{ASSIGN}(q)\|^2\right)$$

$$\leq O\left((2(T+1)wk)^{\frac{1}{1-a-b}}\right) \cdot \Lambda^2 + \sum_{x \in S} O\left(\|x - u^*(x)\|^2 + (r^*(x))^2\right) + \frac{1}{T} \sum_{\substack{i \in [\log \log n] \\ x \in B_i \\ q \in Q(x)}} O\left(\|q - \text{ASSIGN}(q)\|^2\right)$$

$$\leq O\left((2(T+1)wk)^{\frac{1}{1-a-b}}\right) \cdot \Lambda^2 + O(1) \cdot \text{OPT}_S + \frac{1}{T} \sum_{\substack{i \in [\log \log n] \\ x \in B_i \\ q \in Q(x)}} O\left(\|q - \text{ASSIGN}(q)\|^2\right) \quad (2)$$

Now recall that for every $i \in [\log \log n]$ and for every $x \neq y \in B_i$ it holds that $Q(x) \cap Q(y) = \emptyset$. Hence, every point $q \in S$ contributes at most $\log \log n$ times to the last summation above. So,

$$(2) \leq O\left((2(T+1)wk)^{\frac{1}{1-a-b}}\right) \cdot \Lambda^2 + O(1) \cdot \text{OPT}_S + \frac{\log \log n}{T} \sum_{q \in S} O\left(\|q - \text{ASSIGN}(q)\|^2\right)$$

For $T = \Theta(\log \log n)$ (large enough) we get that the last term above is at most half of the left hand side of the inequality, and hence,

$$\text{cost}_S\left(\{\text{ASSIGN}[j] : j \in [k]\}\right) \leq O\left((2(T+1)wk)^{\frac{1}{1-a-b}}\right) \cdot \Lambda^2 + O(1) \cdot \text{OPT}_S$$

$\square$

Combining Lemma 3.9 with Theorem 2.8 yields the following theorem.

**Theorem 3.10.** *There is an $(\varepsilon, \delta)$-differentially private algorithm that, given a database $S$ containing $n$ points in the $d$-dimensional ball $\mathcal{B}(0, \Lambda)$, identifies with probability $1 - \beta$ a $(\gamma, \eta)$-approximation for the $k$-means of $S$, where $\gamma = O(1)$ and $\eta = \text{poly}\left(\log(n), \log(\frac{1}{\beta}), \log(\frac{1}{\delta}), d, \frac{1}{\varepsilon}, k\right) \cdot \Lambda^2$.*

## 4 Private $k$-means – the distributed setting

We begin by describing private computation in the local model where each individual holds her private information locally, and only releases the outcomes of privacy-preserving computations on her data. This is modeled by letting the algorithm access each entry $x_i$ in the input database $S = (x_1, \ldots, x_n) \in X^n$ separately, and only via differentially private *local randomizers*.

**Definition 4.1** (Local Randomizer, LR Oracle [20, 33])**.** *A local randomizer $R : X \to W$ is a randomized algorithm that takes a database of size $n = 1$. Let $S = (x_1, \ldots, x_n) \in X^n$ be a database. An LR oracle $LR_S(i, R)$ gets an index $i \in [n]$ and a local randomizer $R$, and outputs a random value $w \in W$ chosen according to the distribution $R(x_i)$.*

**Definition 4.2** (Local differential privacy [20, 33])**.** *An algorithm satisfies $(\varepsilon, \delta)$-local differential privacy (LDP) if it accesses the database $S$ only via the oracle $LR_S$ with the following restriction: for all possible executions of the algorithm and for all $i \in [n]$, if $LR_S(i, R_1), \ldots, LR_S(i, R_\ell)$ are the algorithm's invocations of $LR_S$ on index $i$, then the algorithm $\mathcal{B}(x) = (R_1(x), R_2(x), \ldots, R_\ell(x))$ is $(\varepsilon, \delta)$-differentially private.*

Local algorithms that prepare all their queries to $LR_S$ before receiving any answers are *non-interactive*; otherwise, they are *interactive*. When $\delta = 0$ we omit it, and say that the algorithm satisfies $\varepsilon$-LDP.

## 4.1 Additional preliminaries

We now present additional preliminaries that enable our construction.

### 4.1.1 Counting queries and histograms with local differential privacy

The most basic task that we can apply in the local differential privacy model is *counting*. Let $S \in \{0,1\}^n$ be a database which is distributed among $n$ users (each holding one bit), and consider the task of estimating the number of users holding a 1. This can be solved privately with error proportional to $\frac{1}{\varepsilon}\sqrt{n}$ (see, e.g., [33]). A more general setting is when instead of a binary domain, every user holds an input item from some (potentially) large domain $X$. This can be solved using tools from the recent line of work on heavy hitters in the local model. [28, 12, 11, 15]

**Notation.** For a database $S = (x_1, \ldots, x_n) \in X^n$ and a domain element $x \in X$, we use $f_S(x)$ to denote the multiplicity of $x$ in $S$, i.e., $f_S(x) = |\{x_i \in S : x_i = x\}|$.

**Theorem 4.3** ([28, 12, 11, 15]). *Fix $\beta, \varepsilon \leq 1$. There exists a non-interactive $\varepsilon$-LDP algorithm that operates on a (distributed) database $S \in Y^n$ for some finite set $Y$, and returns a mapping $\hat{f} : Y \to \mathbb{R}$ such that the following holds. For every choice of $y \in Y$, with probability at least $1 - \beta$, we have that*

$$\left|\hat{f}(y) - f_S(y)\right| \leq O\left(\frac{1}{\varepsilon} \cdot \sqrt{n \cdot \log\left(\frac{1}{\beta}\right)}\right).$$

For our construction, we will need the following extension of Theorem 4.3. This extension is obtained from the analysis of [11] with minor modifications. We include the proof in the appendix for completeness.

**Theorem 4.4** (Algorithm GroupHist). *Fix $\beta, \varepsilon \leq 1$. There exists a non-interactive $\varepsilon$-LDP algorithm that operates on a (distributed) database $S \in Y^n$ for some finite set $Y$, and returns a mapping $\hat{f} : Y \to \mathbb{R}$ such that the following holds. For every choice of a subset $Q \subseteq Y$ with weights $\sigma : Q \to [0,1]$, with probability at least $1 - \beta$, we have that*

$$\left|\sum_{y \in Q} \hat{f}(y) \cdot \sigma(y) - \sum_{y \in Q} f_S(y) \cdot \sigma(y)\right| \leq O\left(\frac{1}{\varepsilon} \cdot \sqrt{|Q| \cdot n \cdot \log\left(\frac{1}{\beta}\right)}\right).$$

### 4.1.2 Algorithm `LDP-GoodCenter`

Let $a > b > 0$ and $c > 1$ be s.t. for every $r > 0$ there exists family $\mathcal{H}$ $(r, cr, p{=}n^{-b}, q{=}n^{-2-a})$-sensitive hash functions mapping $\mathbb{R}^d$ to a universe $U$. The next theorem follows from the results of Nissim and Stemmer [39] with minor modifications, where the constants $a, b, c$ in the theorem depend on the LSH family that the algorithm is instantiated with.

**Theorem 4.5** (Algorithm `LDP-GoodCenter` [39]). *There exists an $\varepsilon$-LDP algorithm that uses a constant number of interactions with the users, such that the following holds. Let $S = (x_1, \ldots, x_n)$ be a database containing $n$ points in the $d$-dimensional ball $\mathcal{B}(0, \Lambda)$, and let $P \subseteq S$ be a fixed subset (unknown to the algorithm) such that for a global constant $\Gamma$*

$$|P| \geq \frac{\Gamma}{\varepsilon} \cdot n^{\frac{2}{3}+a+b} \cdot d^{1/3} \cdot \log(dn) \cdot \log^{1/3}\left(\frac{1}{\beta}\right).$$

*The algorithm outputs a set $Y$ of $O\left(\varepsilon \cdot n^{1/3+a}/\log^{1/3}(n/\beta)\right)$ vectors in $\mathbb{R}^d$ s.t. with probability at least $1-\beta$ there exists a vector $y \in Y$ such that the ball of radius $5c \cdot \operatorname{diam}(P) + \frac{\Lambda}{n}$ around $y$ contains all of $P$.*

## 4.2 An LDP protocol for $k$-means

---
**Algorithm** `LDP-`$k$`-Means`

**Input:** Failure probability $\beta$, privacy parameter $\varepsilon$, and additional LSH parameters $0 < b < a < 1$ and $c = c(a,b) > 1$.

**Setting:** Each player $i \in [n]$ holds a point $x_i$ in the $d$-dimensional ball $\mathcal{B}(0, \Lambda)$. Define $S = (x_1, \ldots, x_n)$.

1. Apply algorithm `LDP-GoodCenter` (Theorem 4.5) on the database $S$ with the privacy parameter $\frac{\varepsilon}{2}$. Obtain a set of $L = O\left(\varepsilon \cdot n^{1/3+a}/\log^{1/3}(n/\beta)\right)$ vectors: $Y = \{y_1, \ldots, y_L\}$.

2. For every $y \in Y$ let $c_y$ denote the number of input points (from $S$) whose closest point in $Y$ is $y$. That is, $c_y = |\{x \in S : y = \operatorname{argmin}_{y' \in Y} \|x - y'\|\}|$. Denote $B = \{(y, c_y) : y \in Y\}$.

3. Apply algorithm `GroupHist` (Theorem 4.4) on the database $S$ with the privacy parameter $\frac{\varepsilon}{2}$ to obtain for every $y \in Y$ an estimation $\hat{c}_y$ for $c_y$.

   % In more details, first every user $i$ (holding an input $x_i \in \mathbb{R}^d$) identifies the center $y_i \in Y$ with minimal distance to $x_i$. This defines a *modified* (distributed) database $S' = (y_1, \ldots, y_n)$. Then, algorithm `GroupHist` is applied to this modified database, and returns a mapping $\hat{f} : Y \to \mathbb{R}$. The weights $\hat{c}_y$ are computed as $\hat{c}_y = \hat{f}(y)$. In the analysis we will use the additional properties of `GroupHist`.

4. Let $\hat{B} = \{(y, \hat{c}_y) : y \in Y\}$.

5. Let $\hat{C}$ be a (non-private) approximation to the $k$-means of $\hat{B}$.

6. Return $\hat{C}$.
---

Similarly to our construction for the centralized model, we will design an algorithm for approximating the $k$-means by first identifying a set of candidate centers, and then choosing a subset of $k$ of them with low $k$-means cost. However, the techniques of Gupta et al. and Balcan et al. (for choosing the subset of $k$ candidate centers) do not directly apply to the local model. Instead, after obtaining the set of candidate centers $Y$, we privately assign (noisy) *weights* to the candidate centers, where the weight of $y \in Y$ is the number of input points whose nearest candidate center is $Y$. We then show that this information (the set of candidate centers with their weights) can be post-processed to obtain an approximation to the $k$-means of the input points.

Consider the execution of algorithm `LDP-`$k$`-Means`, and consider the notations specified in Table 3.

**Claim 4.6.** *For every set of centers $D \subseteq \mathbb{R}^d$ we have $\operatorname{cost}_B(D) \leq 3 \cdot \operatorname{cost}_S(Y_S^*) + 3 \cdot \operatorname{cost}_S(D)$.*

*Proof.* For a point $x$ and a set of centers $C$, let $C(x)$ be the center in $C$ which is closest to $x$. So,

| | |
|---|---|
| $S$ | The input database. |
| $Y$ | The set of $L$ centers chosen on Step 1. |
| $B = \{(y, c_y)\}_{y \in Y}$ | The weighted set of points defined on Step 2. |
| $\hat{B} = \{(y, \hat{c}_y)\}_{y \in Y}$ | The weighted set of points defined on Step 4. |
| $Y_S^* \subseteq Y$ | A subset of $k$ centers from $Y$ that minimizes $\text{cost}_S(\cdot)$. |
| $Y_B^* \subseteq Y$ | A subset of $k$ centers from $Y$ that minimizes $\text{cost}_B(\cdot)$. |
| $Y_{\hat{B}}^* \subseteq Y$ | A subset of $k$ centers from $Y$ that minimizes $\text{cost}_{\hat{B}}(\cdot)$. |

Table 3: Notations for the analysis of algorithm LDP-$k$-Mean

for every set of centers $D \subseteq \mathbb{R}^d$ we have

$$\text{cost}_B(D) = \sum_{y \in Y} c_y \cdot \|y - D(y)\|^2$$
$$= \sum_{x \in S} \|Y(x) - D(Y(x))\|^2$$
$$\leq \sum_{x \in S} \|Y(x) - D(x)\|^2$$
$$\leq \sum_{x \in S} (\|Y(x) - x\| + \|x - D(x)\|)^2$$
$$\leq \sum_{x \in S} \left(3 \cdot \|Y(x) - x\|^2 + 3 \cdot \|x - D(x)\|^2\right)$$
$$\leq 3 \cdot \text{cost}_S(Y_S^*) + 3 \cdot \text{cost}_S(D).$$

□

**Claim 4.7.** *For every set of centers $D \subseteq \mathbb{R}^d$ we have $\text{cost}_S(D) \leq 3 \cdot \text{cost}_S(Y_S^*) + 3 \cdot \text{cost}_B(D)$.*

*Proof.*

$$\text{cost}_S(D) = \sum_{x \in S} \|x - D(x)\|^2$$
$$\leq \sum_{x \in S} \|x - D(Y(x))\|^2$$
$$\leq \sum_{x \in S} (\|x - Y(x)\| + \|Y(x) - D(Y(x))\|)^2$$
$$\leq \sum_{x \in S} \left(3 \cdot \|x - Y(x)\|^2 + 3 \cdot \|Y(x) - D(Y(x))\|^2\right)$$
$$\leq 3 \cdot \text{cost}_S(Y_S^*) + 3 \cdot \text{cost}_B(D).$$

□

**Theorem 4.8.** *Algorithm LDP-$k$-Means satisfies $\varepsilon$-LDP and uses a constant number of interactions with the users. In addition, with probability at least $1 - \beta$ the algorithms returns a $(\gamma, \eta)$-approximation for the k-means of the (distributed) data points, where $\gamma = O(1)$ and $\eta = O\left(\frac{k^2 \cdot \Lambda^2}{\varepsilon} \cdot n^{\frac{2}{3}+a+b} \cdot d^{1/3} \cdot \log(dn) \cdot \log^{1/3}\left(\frac{k}{\beta}\right)\right)$.*

*Proof.* The privacy properties of algorithm `LDP-k-Means` are immediate (follows from composition). We now proceed with the utility analysis. Consider the following event:

> **Event $E_1$ (over the randomness of algorithm `LDP-GoodCenter`):**
>
> $$\mathrm{cost}_S(Y_S^*) \leq O(1) \cdot \mathrm{OPT}_S + O\left(k \cdot t \cdot \Lambda^2\right),$$
>
> where $t \triangleq O\left(\frac{1}{\varepsilon} \cdot n^{\frac{2}{3}+a+b} \cdot d^{1/3} \cdot \log(dn) \cdot \log^{1/3}\left(\frac{k}{\beta}\right)\right)$.

Let $u_1^*, \ldots, u_k^* \in \mathbb{R}^d$ denote an optimal set of centers for $S$, and let $S_j^*$ be the cluster induced by $u_j^*$, i.e., $S_j^* = \{x \in S : j = \mathrm{argmin}_\ell \|x - u_\ell^*\|\}$. For $j \in [k]$ let $r_j^* = \sqrt{\frac{2}{|S_j^*|} \sum_{x \in S_j^*} \|x - u_j^*\|^2}$, and let $P_j^* = \mathcal{B}(u_j^*, r_j^*) \cap S_j^*$.

Fix $j$ such that $|P_j^*| \geq t$. By the properties of algorithm `LDP-GoodCenter` (Theorem 4.5), with probability at least $1 - \frac{\beta}{k}$, there exists a center $y^{(j)} \in Y$ s.t. a ball of radius $O(r_j^* + \frac{\Lambda}{n})$ around $y^{(j)}$ contains all of $P_j^*$. Using the union bound, this is the case for every such $j$ simultaneously with probability at least $1 - \beta$. We continue with the analysis assuming that this is the case. Now, for every $j$ s.t. $|P_j^*| < t$, let $y^{(j)}$ be an arbitrary center in $Y$, and denote $D = \{y^{(1)}, \ldots, y^{(k)}\} \subseteq Y$. We establish an upper bound on the cost of $D$ (w.r.t. the input points $S$) by prescribing a feasible (but not necessarily optimal) assignment of the data points to the centers in $D$: Every point $x \in S$ is assigned to the center $y^{(j)}$, where $j$ is such that $x \in S_j^*$. We have that

$$\mathrm{cost}_S(D) \leq \sum_{j \in [k]} \sum_{x \in S_j^*} \left\|x - y^{(j)}\right\|^2$$

$$\leq k \cdot t \cdot \Lambda^2 + \sum_{\substack{j \in [k]: \\ |P_j^*| \geq t}} \sum_{x \in S_j^*} \left\|x - y^{(j)}\right\|^2$$

$$\leq k \cdot t \cdot \Lambda^2 + \sum_{\substack{j \in [k]: \\ |P_j^*| \geq t}} \sum_{x \in S_j^*} \left(\|x - u_j^*\| + \|u_j^* - y^{(j)}\|\right)^2$$

$$\leq k \cdot t \cdot \Lambda^2 + \sum_{\substack{j \in [k]: \\ |P_j^*| \geq t}} \sum_{x \in S_j^*} \left(\|x - u_j^*\| + O\left(r_j^* + \frac{\Lambda}{n}\right)\right)^2$$

$$\leq k \cdot t \cdot \Lambda^2 + \sum_{\substack{j \in [k]: \\ |P_j^*| \geq t}} \sum_{x \in S_j^*} \left[O\left(\|x - u_j^*\|^2\right) + O\left((r_j^*)^2\right) + O\left(\frac{\Lambda^2}{n^2}\right)\right]$$

$$\leq k \cdot t \cdot \Lambda^2 + \sum_{\substack{j \in [k]: \\ |P_j^*| \geq t}} \sum_{x \in S_j^*} \left[O\left(\|x - u_j^*\|^2\right) + O\left(\frac{1}{|S_j^*|} \sum_{\tilde{x} \in S_j^*} \|\tilde{x} - u_j^*\|^2\right) + O\left(\frac{\Lambda^2}{n^2}\right)\right]$$

$$= k \cdot t \cdot \Lambda^2 + \sum_{\substack{j \in [k]: \\ |P_j^*| \geq t}} \left[ \sum_{x \in S_j^*} O\left( \|x - u_j^*\|^2 \right) + \sum_{x \in S_j^*} O\left( \|x - u_j^*\|^2 \right) \right] + O\left( \frac{\Lambda^2}{n} \right)$$

$$= O\left( k \cdot t \cdot \Lambda^2 \right) + O(1) \cdot \mathrm{OPT}_S.$$

In particular, Event $E_1$ happens with probability at least $1 - \beta$.

> **Event $E_2$ (over the randomness of algorithm `GroupHist`):**
> For every subset $D \subseteq Y$ of size $k$ we have
>
> $$\left| \mathrm{cost}_B(D) - \mathrm{cost}_{\hat{B}}(D) \right| \leq O\left( \frac{k \cdot \Lambda^2}{\varepsilon} \cdot \sqrt{|Y| \cdot n \cdot \log\left( \frac{n}{\beta} \right)} \right).$$

Fix a subset $D = (d_1, \ldots, d_k) \subseteq Y$, and let $G_1^D, \ldots, G_k^D$ denote an optimal partition of the (weighted) points in $Y$, assigning them to the centers in $D$. That is, for every $j \in [k]$ and $y \in G_j^D$ we have that $d_j \in \arg\min_{d \in D} \|y - d\|$. Fix $j \in [k]$. By the properties of algorithm `GroupHist` (Theorem 4.4), with probability at least $1 - \frac{\beta}{k \cdot n^k}$ we have that

$$\left| \sum_{y \in G_j^D} (c_y - \hat{c}_y) \cdot \|y - d_j\|^2 \right| \leq O\left( \frac{\Lambda^2}{\varepsilon} \cdot \sqrt{|G_j^D| \cdot kn \cdot \log\left( \frac{n}{\beta} \right)} \right).$$

Using the union bound, this holds simultaneously for every $j \in [k]$ and every subset $D \subseteq Y$ of size $k$ with probability at least $1 - \beta$. We now show that Event $E_2$ occurs in this case. Indeed, for every such $D \subseteq Y$ we have

$$\mathrm{cost}_B(D) = \sum_{j \in [k]} \sum_{y \in G_j^D} c_y \cdot \|y - d_j\|^2$$

$$= \sum_{j \in [k]} \sum_{y \in G_j^D} \hat{c}_y \cdot \|y - d_j\|^2 + \sum_{j \in [k]} \sum_{y \in G_j^D} (c_y - \hat{c}_y) \cdot \|y - d_j\|^2$$

$$= \mathrm{cost}_{\hat{B}}(D) + \sum_{j \in [k]} \sum_{y \in G_j^D} (c_y - \hat{c}_y) \cdot \|y - d_j\|^2$$

$$\leq \mathrm{cost}_{\hat{B}}(D) + \sum_{j \in [k]} O\left( \frac{\Lambda^2}{\varepsilon} \cdot \sqrt{|G_j^D| \cdot kn \cdot \log\left( \frac{n}{\beta} \right)} \right)$$

$$= \mathrm{cost}_{\hat{B}}(D) + O\left( \frac{\Lambda^2}{\varepsilon} \cdot \sqrt{kn \cdot \log\left( \frac{n}{\beta} \right)} \cdot \sum_{j \in [k]} \sqrt{|G_j^D|} \right)$$

$$\leq \mathrm{cost}_{\hat{B}}(D) + O\left( \frac{k \cdot \Lambda^2}{\varepsilon} \cdot \sqrt{|Y| \cdot n \cdot \log\left( \frac{n}{\beta} \right)} \right),$$

where the last inequality follows from the Cauchy-Schwarz inequality (and by recalling that $\sum_j |G_j^D| = |Y|$). The analysis for the reverse direction is identical. This shows that Event $E_2$ happens with probability at least $1 - \beta$. We continue with the analysis assuming that this is the case.

We are now ready to complete the proof. On Step 5 of algorithm LDP-$k$-Means we identify a non-private approximation $\hat{C}$ for the $k$-means of $\hat{B}$.[4] Together with Event $E_2$ (stating that $\text{cost}_B(D) \approx \text{cost}_{\hat{B}}(D)$ for every subset $D \subseteq Y$ of size $k$) we get that

$$\text{cost}_B\left(\hat{C}\right) \leq O(1) \cdot \text{OPT}_B(Y) + O\left(\frac{k^2 \cdot \Lambda^2}{\varepsilon} \cdot \sqrt{|Y| \cdot n \cdot \log\left(\frac{n}{\beta}\right)}\right)$$

$$= O(1) \cdot \text{cost}_B\left(Y_B^*\right) + O\left(\frac{k^2 \cdot \Lambda^2}{\varepsilon} \cdot \sqrt{|Y| \cdot n \cdot \log\left(\frac{n}{\beta}\right)}\right).$$

Our goal is to bound $\text{cost}_S\left(\hat{C}\right)$. By Claim 4.7,

$$\text{cost}_S\left(\hat{C}\right) \leq 3 \cdot \text{cost}_S\left(Y_S^*\right) + 3 \cdot \text{cost}_B\left(\hat{C}\right)$$

$$\leq 3 \cdot \text{cost}_S\left(Y_S^*\right) + O(1) \cdot \text{cost}_B\left(Y_B^*\right) + O\left(\frac{k^2 \cdot \Lambda^2}{\varepsilon} \cdot \sqrt{|Y| \cdot n \cdot \log\left(\frac{n}{\beta}\right)}\right)$$

$$\leq 3 \cdot \text{cost}_S\left(Y_S^*\right) + O(1) \cdot \text{cost}_B\left(Y_S^*\right) + O\left(\frac{k^2 \cdot \Lambda^2}{\varepsilon} \cdot \sqrt{|Y| \cdot n \cdot \log\left(\frac{n}{\beta}\right)}\right), \quad (3)$$

where the last inequality is because $Y_S^* \subseteq Y$, and $Y_B^*$ minimizes $\text{cost}_B(\cdot)$ over every subset of $Y$ of size $k$. Now, by Claim 4.6 we have that

$$(3) = O(1) \cdot \text{cost}_S\left(Y_S^*\right) + O\left(\frac{k^2 \cdot \Lambda^2}{\varepsilon} \cdot \sqrt{|Y| \cdot n \cdot \log\left(\frac{n}{\beta}\right)}\right) \quad (4)$$

Finally, by Event $E_1$, and be recalling that $|Y| = O\left(\varepsilon \cdot n^{1/3+a} / \log^{1/3}(n/\beta)\right)$, we have

$$(4) \leq O(1) \cdot \text{OPT}_S + O\left(\frac{k^2 \cdot \Lambda^2}{\varepsilon} \cdot n^{\frac{2}{3}+a+b} \cdot d^{1/3} \cdot \log(dn) \cdot \log^{1/3}\left(\frac{k}{\beta}\right)\right).$$

□

## 5 Private coresets for $k$-means

Let $S$ be a set of input points. A *coreset* of $S$ is a small (weighted) set of points $P$ that captures some geometric properties of $S$. In the context of $k$-means, the geometric property that we want

---
[4]A technical issue here is that $\hat{B}$ might contain negative weights. We are not aware of a result for directly approximating the $k$-means of a set of points with negative weights. One option is to replace negative weights with zero, but that would cause our bound on the additive error to increase significantly, roughly from $n^{2/3}$ to $n^{3/4}$ (because negative weights can cancel out other "overly positive" weights). In the appendix we show that the fact that $\text{cost}_{\hat{B}}(\cdot) \approx \text{cost}_B(\cdot)$ can be leveraged to obtain an approximation to the $k$-means of $\hat{B}$ and of $B$. Specifically, we will apply (a variant of) the local search algorithm to the weighted set $\hat{B}$ and analyze the algorithm as if it had been executed on $B$. See the appendix for more details.

$P$ to capture is the $k$-means cost of *every* possible choice for $k$ centers. That is, for every set of $k$ centers $D \subseteq \mathbb{R}^d$ we want that $\text{cost}_P(D) \approx \text{cost}_S(D)$. Formally,

**Definition 5.1** (Coreset for $k$-means)**.** *Let $S$ be a finite set of points in $\mathbb{R}^d$. A finite set of weighted points $P \subseteq (\mathbb{R}^d \times \mathbb{R})$ is an $(\gamma, \eta)$-coreset of $S$ if for every set of $k$ centers $D \subseteq \mathbb{R}^d$ we have*

$$\text{cost}_P(D) \leq \gamma \cdot \text{cost}_S(D) + \eta, \qquad \text{and}$$
$$\text{cost}_S(D) \leq \gamma \cdot \text{cost}_P(D) + \eta.$$

We are interested in *private* algorithms for computing coresets. That is, we are seeking for a differentially private algorithm $\mathcal{A}$ that takes a database $S \in (\mathbb{R}^d)^n$ and outputs a (weighted) set $P$ such that

1. The privacy of the input points in $S$ is preserved; and,

2. The output $P$ allows for approximating the $k$-means cost of every choice for $k$ centers (w.r.t. the input set $S$).

In this section we briefly describe how our techniques from the previous sections can be used to obtain differentially private algorithms for coresets, both for the centralized model and for the local model of differential privacy.

Consider again Algorithm `LDP-k-Means`. Given an input set $S$, we privately identified a set $Y$ of candidate centers, and privately estimated the *weights* of every center $y \in Y$, where the *weight* of a candidate center $y$ is the number of input points that $y$ is their closest neighbor in $Y$. We denoted the resulting assignment of (noisy) weights to the candidate centers as $\hat{B}$, and used $B$ to denote these points with their "true" weights. In the analysis, we then argued that for every set of $k$ centers $D \subseteq Y$ we have

$$\text{cost}_S(D) \approx \text{cost}_B(D) \approx \text{cost}_{\hat{B}}(D). \tag{5}$$

With the coreset objective in mind, the above equation suggests that our (privately computed) $\hat{B}$ might actually be a coreset of $S$. However, Definition 5.1 requires our approximation guarantees to hold for every choice of $k$ centers from $\mathbb{R}^d$, whereas Equation (5) only holds for every choice of $k$ centers out of $Y$ (our precomputed set of candidate centers). Actually, the approximation $\text{cost}_S(D) \approx \text{cost}_B(D)$ *does* hold for every $D \subseteq \mathbb{R}^d$ (see Claims 4.6 and 4.7), and it is only the connection between $\text{cost}_B(D)$ and $\text{cost}_{\hat{B}}(D)$ that requires attention. Specifically, in the analysis of Event $E_2$, we first showed that for every fixed set of centers $D$ we have that $\text{cost}_B(D) \approx \text{cost}_{\hat{B}}(D)$, and then we used the union bound over every possible choice for $k$ centers out of $Y$. As $|Y| = \text{poly}(n)$, this increased our error by a factor of at most $\sqrt{\log(n^k)}$, which was acceptable. However, this argument fails if the centers in $D$ come from an infinite domain.

To recover from this difficulty and get a coreset after all, we add a step to the algorithm in which we we take the set $\hat{C}$ of $k$-centers that approximates the optimal $k$-means (computed in Step 5), and recompute approximate weights $\hat{c}_z$ for every $z \in \hat{C}$. That is, we rerun Step 3 of Algorithm `LDP-k-Means` with $\hat{C}$ as the set $Y$ and return the (weighted) set $\hat{C}$ as our coreset.

Every possible choice $D$ of $k$ centers in $\mathbb{R}^d$ induce a partition of $\hat{C}$ in which each part consists of the centers in $\hat{C}$ that are closest to a particular element of $D$. So it suffices to apply the union bound to every possible partitioning of $\hat{C}$ into $k$ groups and argue that the weighted cost of such a partition (with respect to any set of centers $D$ that induce it) is close to the cost of the entire data set with respect to $D$. We obtain the following theorems:

**Theorem 5.2.** *There is an $(\varepsilon, \delta)$-differentially private algorithm that, given a database $S$ containing $n$ points in the $d$-dimensional ball $\mathcal{B}(0, \Lambda)$, identifies with probability $1 - \beta$ a $(\gamma, \eta)$-coreset of $S$, where $\gamma = O(1)$ and $\eta = \text{poly}\left(\log(n), \log(\frac{1}{\beta}), \log(\frac{1}{\delta}), d, \frac{1}{\varepsilon}, k\right) \cdot \Lambda^2$.*

**Theorem 5.3.** *There is an $\varepsilon$-LDP algorithm that uses a constant number of interactions with the users. In addition, with probability at least $1 - \beta$ the algorithms returns a $(\gamma, \eta)$-coreset for the (distributed) data points, where $\gamma = O(1)$ and $\eta = \text{poly}\left(\log(\frac{1}{\beta}), d, \frac{1}{\varepsilon}, k\right) \cdot n^{0.67} \cdot \Lambda^2$.*

**Acknowledgments.** We thank Moni Naor for helpful discussions of ideas in this work.

[38] K. Nissim, S. Raskhodnikova, and A. Smith. Smooth sensitivity and sampling in private data analysis. In *STOC*, pages 75–84. ACM, 2007.

[39] K. Nissim and U. Stemmer. Clustering algorithms for the centralized and local models. In F. Janoos, M. Mohri, and K. Sridharan, editors, *Proceedings of Algorithmic Learning Theory*, volume 83 of *Proceedings of Machine Learning Research*, pages 619–653. PMLR, 07–09 Apr 2018.

[40] K. Nissim, U. Stemmer, and S. P. Vadhan. Locating a small cluster privately. In *Proceedings of the 35th ACM SIGMOD-SIGACT-SIGAI Symposium on Principles of Database Systems, PODS 2016, San Francisco, CA, USA, June 26 - July 01, 2016*, pages 413–427, 2016.

[41] R. Nock, R. Canyasse, R. Boreli, and F. Nielsen. k-variates++: more pluses in the k-means++. In *Proceedings of the 33nd International Conference on Machine Learning, ICML 2016, New York City, NY, USA, June 19-24, 2016*, pages 145–154, 2016.

[42] J. P. Schmidt, A. Siegel, and A. Srinivasan. Chernoff-Hoeffding bounds for applications with limited independence. *SIAM Journal on Discrete Mathematics*, 8(2):223–250, 1995.

[43] D. Su, J. Cao, N. Li, E. Bertino, and H. Jin. Differentially private k-means clustering. In *Proceedings of the Sixth ACM Conference on Data and Application Security and Privacy*, CODASPY '16, pages 26–37, New York, NY, USA, 2016. ACM.

[44] S. Vadhan. *The Complexity of Differential Privacy*. 2016.

[45] Y. Wang, Y.-X. Wang, and A. Singh. Differentially private subspace clustering. In *Proceedings of the 28th International Conference on Neural Information Processing Systems - Volume 1*, NIPS'15, pages 1000–1008, Cambridge, MA, USA, 2015. MIT Press.

## A  Proof of Lemma 3.1

The privacy properties of `LSH-Procedure` are straight forward (follow from composition). We now proceed with the utility analysis. Let $P \subseteq S$ be s.t. $|P| = t$ and $\mathrm{diam}(P) \leq r$, and consider the following good event.

> **Event $E_1$ (over partitioning $S$ into $S_1, \ldots, S_M$):**
> For every $m \in [M]$ we have $|P \cap S_m| \geq \frac{t}{2M}$.

As in Theorem 2.10 (the Poisson approximation), we analyze event $E_1$ in the Poisson case. To that end, let $J_1, \cdots, J_M$ be independent Poisson random variables with mean $\frac{t}{M}$. Let us say that $m \in [M]$ is *bad* if $J_m < \frac{t}{2M}$. Now fix $m \in [M]$. Using a tail bound for the Poisson distribution (see Theorem 2.11), assuming that $t \geq 24M \ln(\frac{eM}{\beta})$, we have that $m$ is bad with probability at most $\frac{\beta}{eM\sqrt{t}}$. By a union bound, the probability that a bad $m$ exists is at most $\frac{\beta}{e\sqrt{t}}$. Hence, by the Poisson approximation, we get that $\Pr[E_1] \geq 1 - \beta$. We proceed with the analysis assuming that Event $E_1$ occurred.

Fix again $m \in [M]$. By the properties of the family $\mathcal{H}$, for every $x, y \in \mathbb{R}^d$ s.t. $\|x - y\| \geq cr$ we have that $\Pr_{h_m \in \mathcal{H}}[h_m(x) = h_m(y)] \leq q = n^{-2-a}$. Using the union bound we get

$$\Pr_{h_m \in_R \mathcal{H}}[h_m(x) \neq h_m(y) \text{ for all } x, y \in S_m \text{ s.t. } \|x - y\| \geq cr] \geq (1 - n^{-a}/2).$$

Let $x$ be an arbitrary point in $P \cap S_m$. By linearity of expectation, we have that

$$\mathbb{E}_{h_m \in \mathcal{H}}[|\{y \in P \cap S_m : h_m(y) \neq h_m(x)\}|] \leq |P \cap S_m| \cdot (1 - p) = |P \cap S_m| \cdot (1 - n^{-b}).$$

Hence, by Markov's inequality,

$$\Pr_{h_m \in \mathcal{H}}\left[|\{y \in P \cap S_m : h_m(y) \neq h_m(x)\}| \geq \frac{|P \cap S_m| \cdot (1 - n^{-b})}{1 - n^{-a}}\right] \leq 1 - n^{-a}.$$

So,

$$\Pr_{h_m \in \mathcal{H}}\left[|\{y \in P \cap S_m : h_m(y) = h_m(x)\}| \geq |P \cap S_m| \cdot \left(1 - \frac{1 - n^{-b}}{1 - n^{-a}}\right)\right] \geq n^{-a}.$$

Simplifying, for large enough $n$ (specifically, for $n^{a-b} \geq 2$) we get

$$\Pr_{h_m \in \mathcal{H}}\left[|\{y \in P \cap S_m : h_m(y) = h_m(x)\}| \geq |P \cap S_m| \cdot \frac{n^{-b}}{2}\right] \geq n^{-a}.$$

As Event $E_1$ has occurred, we get

$$\Pr_{h_m \in \mathcal{H}}\left[|\{y \in P \cap S_m : h_m(y) = h_m(x)\}| \geq \frac{t \cdot n^{-b}}{4M}\right] \geq n^{-a}.$$

We have established that with probability at least $n^{-a}/2$ over the choice of $h_m \in \mathcal{H}$ in Step 2 the following events occur:

($E_2$) For every $x, y \in S_m$ s.t. $\|x - y\| \geq cr$ it holds that $h_m(x) \neq h_m(y)$; and,

($E_3$) At least $\frac{t \cdot n^{-b}}{4M}$ points from $P \cap S_m$ are mapped into the same value by $h_m$.

Now recall that on Step 2 we choose $M$ independent hash functions $(h_1, \ldots, h_M)$. Hence, for $M \geq 2n^a \ln(\frac{1}{\beta})$, with probability at least $1 - \beta$ there exists $m^* \in [M]$ for which events $E_2, E_3$ occur. In such a case, we get that by $E_3$ there is a hash value $u^*$ s.t.

$$|S_{m^*, u^*}| \geq |\{y \in P \cap S_{m^*} : h_{m^*}(y) = u^*\}| \geq \frac{t \cdot n^{-b}}{4M},$$

and furthermore, by $E_2$, for every $x, y \in S_m$ s.t. $h_{m^*}(x) = h_{m^*}(y) = u^*$ we have that $\|x - y\| \leq cr$. We continue with the analysis assuming that this is the case.

In Step 3 we use the Laplace mechanism to obtain estimations $\hat{w}_{m,u} \approx |S_{m,u}|$ for every $(m, u) \in [M] \times U$, and to construct a list $L \subseteq [M] \times U$ that contains all pairs $(m, u)$ with large estimations. Recall that, w.l.o.g., we have that $|U| \leq n^3$. Hence, by the properties of the Laplace mechanism (Theorem 2.5), with probability at least $1 - \beta$, all of our estimations are accurate to within $\frac{30}{\varepsilon} \ln(\frac{n}{\beta})$. Thus, assuming that $\frac{t \cdot n^{-b}}{4M} \geq \frac{90}{\varepsilon} \ln(\frac{n}{\beta})$, we have that $(m^*, u^*) \in L$. (As a side note, observe that there could be at most $\varepsilon n / \ln(\frac{n}{\beta})$ elements in $L$, which bounds the size of the output.)

Denote the average of the points in $S_{m^*,u^*}$ as $y^*$, and observe that by $E_2$, a ball of radius $cr$ around $y^*$ contains all of $S_{m^*,u^*}$. In Step 4 we use the algorithm from Theorem 2.7 to compute the noisy average $\hat{y}_{m^*,u^*}$ of $S_{m^*,u^*}$. The noise magnitude reduces with the size of the set $|S_{m^*,u^*}|$ (recall that $|S_{m^*,u^*}| \geq \frac{t \cdot n^{-b}}{4M}$), and for $t \geq 4Mn^b \cdot \frac{1152\sqrt{d}}{\varepsilon/2} \ln\left(\frac{4n}{\beta}\right) \sqrt{\ln\left(\frac{16}{\delta}\right)}$, with probability at least $(1-\beta)$ we have that $\|y^* - \hat{y}_{m^*,u^*}\| \leq cr$. In such a case we have that a ball of radius $2cr$ around $\hat{y}_{m^*,u^*}$ contains all of $S_{m^*,u^*}$, and in particular, contains some of the points from $P$. Hence, as $P$ is of diameter $r$, we get that a ball of radius $(2c+1)r$ around $\hat{y}_{m^*,u^*}$ contains *all* of $P$.

All in all, with probability at least $1 - 4\beta$ we have that the output on Step 5 is a set of centers $\{\hat{y}_{m,u} : u \in L\}$ such that a ball of radius $(2c+1)r$ around one of these centers contains all of $P$.

## B  Algorithm GroupHist

In this section we present the proof of Theorem 4.4. The analysis is almost identical to the analysis of [11]. We include the details here for completeness. We first restate Theorem 4.4.

**Theorem B.1** (Algorithm GroupHist). *Fix $\beta, \varepsilon \leq 1$. There exists a non-interactive $\varepsilon$-LDP algorithm that operates on a (distributed) database $S \in Y^n$ for a finite set $Y$ and returns a mapping $\hat{f} : Y \to \mathbb{R}$ such that the following holds. For every choice of a subset $Q \subseteq Y$ with weights $\sigma : Q \to [0,1]$, with probability at least $1 - \beta$, we have that*

$$\left| \sum_{y \in Q} \hat{f}(y) \cdot \sigma(y) - \sum_{y \in Q} f_S(y) \cdot \sigma(y) \right| \leq O\left( \frac{1}{\varepsilon} \cdot \sqrt{|Q| \cdot n \cdot \log\left(\frac{1}{\beta}\right)} \right).$$

The protocol, GroupHist, uses the following simple local randomizer $\mathcal{R}$.

---
**Algorithm $\mathcal{R}$: Basic Randomizer**

**Inputs:** $x \in \{\pm 1\}$, and privacy parameter $\varepsilon$.

1. Generate and return a random bit $y = \begin{cases} x & \text{w.p. } e^\varepsilon/(e^\varepsilon + 1) \\ -x & \text{w.p. } 1/(e^\varepsilon + 1) \end{cases}$

---

**Algorithm GroupHist**

**Public randomness:** Uniformly random matrix $Z \in \{\pm 1\}^{s \times n}$, where $s = |Y|$. Each row of $Z$ is accessed by an element $y \in Y$.

**Setting:** Each player $i \in [n]$ holds a private value $y_i \in Y$. Define $S = (y_1, \cdots, y_n)$.
Define $\widetilde{S} = (\tilde{y}_1, \cdots, \tilde{y}_n)$ where $\tilde{y}_i = Z[y_i, i]$.

1. For $i \in [n]$ let $z_i \leftarrow LR_{\widetilde{S}}(i, \mathcal{R})$.

2. For every $y \in Y$, define $\hat{f}(y) = \frac{e^\varepsilon + 1}{e^\varepsilon - 1} \cdot \sum_{i \in [n]} z_i \cdot Z[y, i]$.

3. Output $\hat{f}(\cdot)$.

---

*Proof.* Fix $Q \subseteq Y$ and $\sigma : Q \to [0,1]$. We start by analyzing the expectation of $\sum_{y \in Q} \sigma(y) \cdot \hat{f}(y)$:

$$\mathbb{E}\left[\sum_{y \in Q} \sigma(y) \cdot \hat{f}(y)\right] = \mathbb{E}\left[\sum_{y \in Q} \sigma(y) \cdot \left(\frac{e^\varepsilon + 1}{e^\varepsilon - 1} \cdot \sum_{i \in [n]} y_i \cdot Z[y, i]\right)\right]$$

$$= \frac{e^\varepsilon + 1}{e^\varepsilon - 1} \cdot \sum_{y \in Q} \sigma(y) \cdot \sum_{i \in [n]} \mathbb{E}\left[z_i \cdot Z[y, i]\right]$$

$$= \frac{e^\varepsilon + 1}{e^\varepsilon - 1} \cdot \sum_{y \in Q} \sigma(y) \cdot \left(\sum_{i \in [n] : y_i = y} \mathbb{E}\left[z_i \cdot Z[y, i]\right] + \sum_{i \in [n] : y_i \neq y} \mathbb{E}\left[z_i \cdot Z[y, i]\right]\right)$$

$$= \frac{e^\varepsilon + 1}{e^\varepsilon - 1} \cdot \sum_{y \in Q} \sigma(y) \cdot \left(\sum_{i \in [n] : y_i = y} \frac{e^\varepsilon - 1}{e^\varepsilon + 1} + \sum_{i \in [n] : y_i \neq y} 0\right)$$

$$= \sum_{y \in Q} \sigma(y) \cdot f_S(y).$$

That is, $\sum_{y \in Q} \sigma(y) \cdot \hat{f}(y)$ can be expressed as the sum of $|Q| \cdot n$ independent random variables with expectation $\sum_{y \in Q} \sigma(y) \cdot f_S(y)$. The theorem now follows from the Hoeffding bound. □

**Remark B.2.** *For the analysis above it suffices that the entries $Z$ are only $k$-wise independent, for $k = O\left(\log(\frac{1}{\beta})\right)$. See, e.g., [42] for a statement of the Hoeffding bound for variables with limited independence.*

## C  Non-private approximation for $k$-means clustering [32, 25, 8]

In Section 4 we used a non-private algorithm for approximating the $k$-means of a weighted set of points $\hat{B}$, with possibly negative weights. We are not aware of a result for directly approximating the $k$-means of a set of points with negative weights. However, in our case the set $\hat{B}$ is not arbitrary, in the sense that $\hat{B}$ is "close" to a set of points $B$ with non-negative weights. This fact can be leveraged to obtain an approximation to the $k$-means of $\hat{B}$, e.g., by modifying the algorithm of [32] using the techniques of [25, 8]. We include their analysis here for completeness. We remark that the presentation here does not attempt to optimize constants, and refer the reader to [32, 25, 8] for a more detailed account.

**Lemma C.1** ([32]). *Let $S \in (\mathbb{R}^d)^n$ be a set of points, let $Y \subseteq \mathbb{R}^d$ be a set of centers, and let $D \subseteq Y$ be a subset of $k$ centers. There exists $x \in D$ and $y \in Y$ such that*

$$\text{cost}_S(D) - \text{cost}_S(D \setminus \{x\} \cup \{y\}) \geq \frac{1}{2k} \left(\text{cost}_S(D) - 256 \cdot \text{OPT}_S(Y)\right).$$

*Proof.* Let $D \subseteq Y$ be a subset of $k$ centers, refereed to as *heuristic* centers. Let $Y^* \subseteq Y$ denote a set of $k$ optimal centers, s.t. $\text{cost}_S(Y^*) = \text{OPT}_S(Y)$. For each optimal center $o \in Y^*$, let $d_o \in D$ denote is closest heuristic center. We say that $o$ is *captured* by $d_o$. Note that each optimal center is captured by exactly one heuristic center, but each heuristic center may capture any number of optimal centers. We say that a heuristic center is *lonely* if it captures no optimal center. The analysis is based on constructing a set of *swap pairs*.

We begin by defining a simultaneous partition of the heuristic centers and the optimal centers into two sets of groups $D_1, D_2, \ldots, D_r$ and $Y_1^*, Y_2^*, \ldots, Y_r^*$ for some $r$, such that $|D_i| = |Y_i^*|$ for all $i$. For each heuristic center $d$ that captures some number $m \geq 1$ of optimal centers, we form a group of $m$ optimal centers consisting if these captured centers. The corresponding group of heuristic centers consists of $d$ together with any $m-1$ lonely heuristic centers. (See [32] for an illustration and for more details.)

We generate the swap pairs as follows. For every partition that involves one captured center we generate a swap pair consisting of the heuristic center and its captured center. For every partition containing two or more captured centers we generate swap pairs between the lonely heuristic centers and the optimal centers, so that each optimal center is involved in exactly one swap pair and each lonely heuristic center is involved in at most two swap pairs. It is easy to verify that:

(1) each optimal center is swapped in exactly once,

(2) each heuristic center is swapped out at most twice, and

(3) if $d$ and $o$ are swapped, then $d$ does not capture any optimal center other than $o$.

**Notations.** For a set of points $S$, a set of centers $Z$, and a center $z \in Z$, we write $N_S(z, Z)$ to denote the set of points from $S$ that are closer to $z$ than to any other center in $Z$. For a set of points $S$, a set of centers $Z$, and a point $x \in S$, we write $Z(x)$ to denote the nearest neighbor of $x$ in $Z$.

For each swap pair $\langle d, o \rangle$ we establish an upper bound on the difference $\bigl(\text{cost}_S(D \setminus \{d\} \cup \{o\}) - \text{cost}_S(D)\bigr)$ by prescribing a feasible (but not necessarily optimal) assignment of the data points to the centers $D \setminus \{d\} \cup \{o\}$. First, the data points in $N_S(o, Y^*)$ are assigned to $o$, implying a change of

$$\sum_{x \in N_S(o, Y^*)} \left( \|x - o\|^2 - \|x - D(x)\|^2 \right). \tag{6}$$

Each point $x \in N_S(d, D) \setminus N_S(o, Y^*)$ has lost $d$ as a center and must be *reassigned* to a new center. Since $x$ is not in $N_S(o, Y^*)$ we know that $Y^*(x) \neq o$, and hence by property (3) above $d$ does not capture $Y^*(x)$. Therefore $D(Y^*(x))$, the nearest heuristic center to $Y^*(x)$, exists after the swap. We assign $x$ to $D(Y^*(x))$. Thus the change due to this reassignment is at most

$$\sum_{x \in N_S(d, D) \setminus N_S(o, Y^*)} \left( \|x - D(Y^*(x))\|^2 - \|x - d\|^2 \right). \tag{7}$$

So, by (6) and (7) we have that

$$\text{cost}_S(D \setminus \{d\} \cup \{o\}) - \text{cost}_S(D) \leq \sum_{x \in N_S(o, Y^*)} \left( \|x - o\|^2 - \|x - D(x)\|^2 \right)$$

$$+ \sum_{x \in N_S(d, D) \setminus N_S(o, Y^*)} \left( \|x - D(Y^*(x))\|^2 - \|x - d\|^2 \right). \tag{8}$$

We now want to sum Inequality 8 over over all swap pairs. To that end, recall that each optimal center is swapped in exactly once, and hence each point $x$ contributes once to the first sum. Also note that the quantity in the second sum is always non-negative, because $D(Y^*(x)) \in D$ and $d$ is the closest center in $D$ to $x$. Hence, by extending the sum to all $x \in N_S(d, D)$ we can only increase its value. Recalling that each heuristic center is swapped in at most twice we have

$$\sum_{\langle d,o \rangle} \big(\text{cost}_S(D \setminus \{d\} \cup \{o\}) - \text{cost}_S(D)\big)$$

$$\leq \sum_{x \in S} \big(\|x - Y^*(x)\|^2 - \|x - D(x)\|^2\big) + 2\sum_{x \in S} \big(\|x - D(Y^*(x))\|^2 - \|x - D(x)\|^2\big)$$

$$= \sum_{x \in S} \|x - Y^*(x)\|^2 - 3\sum_{x \in S} \|x - D(x)\|^2 + 2\sum_{x \in S} \|x - D(Y^*(x))\|^2$$

$$= \text{cost}_S(Y^*) - 3 \cdot \text{cost}_S(D) + 2\sum_{x \in S} \|x - D(Y^*(x))\|^2$$

$$\leq \text{cost}_S(Y^*) - 3 \cdot \text{cost}_S(D) + 2\sum_{x \in S} \big(\|x - Y^*(x)\| + \|Y^*(x) - D(Y^*(x))\|\big)^2$$

$$\leq \text{cost}_S(Y^*) - 3 \cdot \text{cost}_S(D) + 2\sum_{x \in S} \big(\|x - Y^*(x)\| + \|Y^*(x) - D(x)\|\big)^2$$

$$\leq \text{cost}_S(Y^*) - 3 \cdot \text{cost}_S(D) + 2\sum_{x \in S} \big(\|x - Y^*(x)\| + \|x - Y^*(x)\| + \|x - D(x)\|\big)^2$$

$$= \text{cost}_S(Y^*) - 3 \cdot \text{cost}_S(D) + 2\sum_{x \in S} \big(4 \cdot \|x - Y^*(x)\|^2 + 4 \cdot \|x - Y^*(x)\| \cdot \|x - D(x)\| + \|x - D(x)\|^2\big)$$

$$= 9 \cdot \text{cost}_S(Y^*) - \text{cost}_S(D) + 8\sum_{x \in S} \|x - Y^*(x)\| \cdot \|x - D(x)\|$$

$$\leq 9 \cdot \text{cost}_S(Y^*) - \text{cost}_S(D) + 8\sqrt{\text{cost}_S(Y^*) \cdot \text{cost}_S(D)}$$

$$\leq 9 \cdot \text{cost}_S(Y^*) - \frac{1}{2} \cdot \text{cost}_S(D),$$

where for the last inequality we assumed that $\text{cost}_S(D) \geq 256 \cdot \text{cost}_S(Y^*)$. So,

$$\sum_{\langle d,o \rangle} \big(\text{cost}_S(D) - \text{cost}_S(D \setminus \{d\} \cup \{o\})\big) \geq \frac{1}{2} \cdot \text{cost}_S(D) - 9 \cdot \text{cost}_S(Y^*),$$

and hence, there must exist a swap pair $\langle d, o \rangle$ such that

$$\text{cost}_S(D) - \text{cost}_S(D \setminus \{d\} \cup \{o\}) \geq \frac{1}{2k} \cdot \text{cost}_S(D) - \frac{9}{k} \cdot \text{cost}_S(Y^*).$$

$\square$

**Theorem C.2** ([25, 8]). *Let $S \in (\mathbb{R}^d)^n$ be a set of points such that $\text{diam}(S) \leq \Lambda$, and let $Y \subseteq \mathbb{R}^d$ be a set of centers. Let $f : Y^k \to \mathbb{R}$ be a function that given a set of $k$ centers $D \in (\mathbb{R}^d)^k$ returns an arbitrary value such that $|f(D) - \text{cost}_S(D)| \leq \Delta$. There exists an algorithm that identifies a set of $k$ centers $C \subseteq Y$ such that $\text{cost}_S(C) \leq O(1) \cdot \text{OPT}_S(Y) + O(\Lambda^2 + k\Delta)$, using only oracle access to the function $f$ (and no other access to $S$).*

**Algorithm** `NoisyLocalSearch` [32, 25, 8]

**Settings:** Let $Y \subseteq \mathbb{R}^d$ be a (known) set of centers, and let $S \in (\mathbb{R}^d)^n$ be a (fixed but unknown) set of points. The algorithm has oracle access to a function $f : Y^k \to \mathbb{R}$, where for every subset $D \subseteq Y$ of size $k$, the function returns an arbitrary number such that $|f(D) - \text{cost}_S(D)| \leq \Delta$.

1. Arbitrarily select a subset $D^{(0)} \subseteq Y$ of size $k$.
2. Denote $T = 2k \log n$.
3. For $t = 1, 2, \ldots, T$: Choose $(x, y) \in D^{(t-1)} \times Y$ maximizing $f\left(D^{(t-1)}\right) - f\left(D^{(t-1)} \setminus \{x\} \cup \{y\}\right)$, and set $D^{(t)} \leftarrow D^{(t-1)} \setminus \{x\} \cup \{y\}$.
4. Output $Z^{(T)}$.

---

*Proof.* The proof is via the construction of algorithm `NoisyLocalSearch`. By our assumption on the function $f$ and by Lemma C.1, every iteration $t$ identifies a swap pair $(x, y)$ such that

$$\text{cost}_S\left(D^{(t-1)}\right) - \text{cost}_S\left(D^{(t)}\right) = \text{cost}_S\left(D^{(t-1)}\right) - \text{cost}_S\left(D^{(t-1)} \setminus \{x\} \cup \{y\}\right)$$
$$\geq \frac{1}{2k} \cdot \text{cost}_S\left(D^{(t-1)}\right) - \frac{128}{k} \cdot \text{OPT}_S(Y) - 4\Delta.$$

Hence, after $T = 2k \log n$ iterations we have that

$$\text{cost}_S\left(D^{(T)}\right) \leq \left(1 - \frac{1}{2k}\right)^T \cdot \text{cost}_S\left(D^{(0)}\right) + \left(\frac{128}{k} \cdot \text{OPT}_S(Y) + 4\Delta\right) \cdot \sum_{t=0}^{T-1}\left(1 - \frac{1}{2k}\right)^t$$
$$\leq \left(1 - \frac{1}{2k}\right)^T \cdot n \cdot \Lambda^2 + \left(\frac{128}{k} \cdot \text{OPT}_S(Y) + 4\Delta\right) \cdot 2k$$
$$\leq \Lambda^2 + 256 \cdot \text{OPT}_S(Y) + 8k\Delta.$$

□